 \documentclass[preprint,review,12pt]{elsarticle}

\usepackage{graphicx}
\usepackage{amssymb}
\usepackage{algorithmic}
\usepackage[linesnumbered,boxed]{algorithm2e}
\usepackage[caption=false,font=normalsize,labelfont=sf,textfont=sf]{subfig}
\usepackage{url}

\SetKwInOut{KwInput}{Input}
\SetKwInOut{KwOutput}{Output}
\usepackage{float}
\newcommand{\Eqn}[1]{(#1)}
\newcommand{\unit}[0]{mL/min/1.73$m^2$}

\journal{Journal of Biomedical Informatics}

\begin{document}

\begin{frontmatter}


\title{Probabilistic Broken-Stick Model: A Regression Algorithm for Irregularly Sampled Data with Application to eGFR}


\author[label1]{Norman Poh}
\author[label1]{Simon Bull}
\author[label1]{Santosh Tirunagari}
\author[label3]{Nicholas Cole}
\author[label3]{Simon de Lusignan}
\address[label1]{Department of Computer Science, University of Surrey}
%\address[label2]{South West Thames Renal Department, Epsom and St Helier NHS Trust}
\address[label3]{Department of Clinical and Experimental Medicine, University of Surrey}

%\address{California, United States}

\begin{abstract}

In order for clinicians to manage disease progression and make effective decisions about drug dosage, treatment regimens or scheduling follow up appointments, it is necessary to be able to identify both short and long-term trends in repeated biomedical measurements. However, this is complicated by the fact that these measurements are irregularly sampled and influenced by both genuine physiological changes and external factors. In their current forms, existing regression algorithms often do not fulfil all of a clinician's requirements for identifying short-term events while still being able to identify long-term trends in disease progression. Therefore, in order to balance both short term interpretability and long term flexibility, an extension to broken-stick regression models is proposed in order to make them more suitable for modelling clinical time series. The proposed probabilistic broken-stick model can robustly estimate both short-term and long-term trends simultaneously, while also accommodating the unequal length and irregularly sampled nature of clinical time series. Moreover, since the model is parametric and completely generative, its first derivative provides a long-term non-linear estimate of the annual rate of change in the measurements more reliably than linear regression. The benefits of the proposed model are illustrated using estimated glomerular filtration rate as a case study for managing patients with chronic kidney disease.

\end{abstract}


\end{frontmatter}

%\linenumbers

\section{Introduction}

The trend of measurements of clinical interest such as blood sugar, cholesterol or kidney function can provide insight into the expected development of a patient's condition. For patients with chronic illnesses such as diabetes and chronic kidney disease (CKD), monitoring of these measurements is necessary in order to effectively manage the condition. For example, in order for clinicians to make effective decisions about drug dosage, treatment regimens or when schedule follow up appointments it is necessary to know not only the value of these indicators, but also to have an idea of both the short- and long-term trajectory they are following. However, modelling the trend of biomedical measurements over the long-term can be complicated by both practical, e.g. the irregular taking of measurements and lengthy gaps between them, and biological considerations. For example, the primary indicator of kidney function, the estimated glomerular filtration rate (eGFR), can be influenced by, amongst other things, the level of protein in the diet, changes in muscle breakdown and the level of hydration~\cite{de2010creatinine}. This can lead to substantive variability in a patient's eGFR measurements~\cite{poh2012calibrating, poh2014challenges}. Unfortunately, existing regression algorithms such as linear, polynomial and Gaussian process regression (GPR)~\cite{williams2006gaussian} either can not account for these challenges or do not satisfy the key clinical requirements of providing an easily interpretable model that can elucidate short- and long-term trends.

\begin{figure}[ht]
\centering
\begin{tabular}{cc}
\includegraphics[scale=0.35]{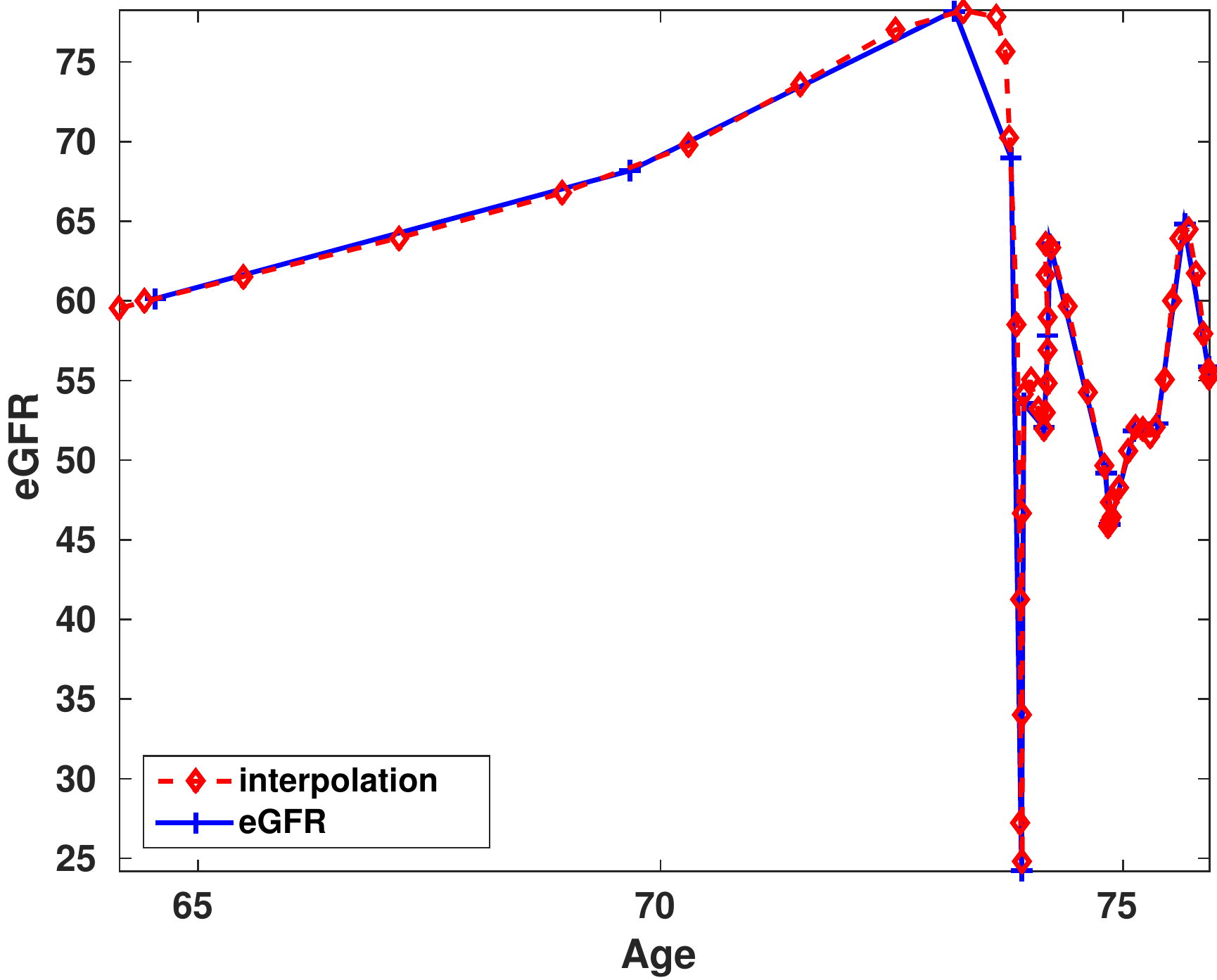}\\
\end{tabular}
\caption{An eGFR time series (blue) modelled using linear interpolation in order to produce a fixed-size vector of $50$ observations (red) over the age range for which the patient has eGFR measurements~\cite{tirunagari2016automatic}.} 
\label{interpolation}
\end{figure}

As biomedical measurements are irregularly sampled, they pose an additional difficulty due to the prior work in time series analysis primarily focussing on regularly sampled data. Despite methods for anaysing irregular time series data directly having been employed successfully~\cite{schulz1997spectrum,stoica2011new,rehfeld2011comparison}, the most common approach is still to transform the data to enforce regularity using either interpolation techniques or regression analysis~\cite{kreindler2006effects}. However, with biomedical time series interpolation can present its own problems due to the measurements not always being taken at random, but rather requested at specific times by clinicians, e.g. as part of routine monitoring or as follow up to treatment. On the other hand, regression imposes a number of assumptions on both the variables and their relationships. For example, linear regression assumes a linear relationships between the dependent and independent variables and independence of the residuals (no auto-correlation); assumptions which are usually violated in biomedical time series. Often linearity is violated due to an acute episode. For example, when a patient suffers an acute kidney injury (AKI)~\cite{tirunagari2016Detection, Druml2014, Faubel2016, Shiao2015} their eGFR will drop sharply and potentially recover a short time after (as seen in Figure~\ref{interpolation}). Long-term trends may therefore exhibit local fluctuations due to genuine physiological changes as well as external factors.

%\begin{figure}[ht]
%\centering
%\begin{tabular}{cc}
%\includegraphics[scale=0.35]{Pictures_B/Bint2.eps} &\includegraphics[scale=0.35]{Pictures_B/problem.eps} \\
%(a) & (b) \\
%\end{tabular}
%\caption{(a) eGFR series (blue) modelled using linear interpolation to produce a fixed-size vector of $50$ observations (red) over the range for which a patient has eGFR measurements. (b) A patient's eGFR is observed at irregular time intervals and over different age ranges. Each colour represents a single patient. The images are obtained from~\cite{tirunagari2016automatic}.} 
%\label{interpolation}
%\end{figure}

\begin{figure}[ht]
\centering
\begin{tabular}{c}
\includegraphics[scale=0.35]{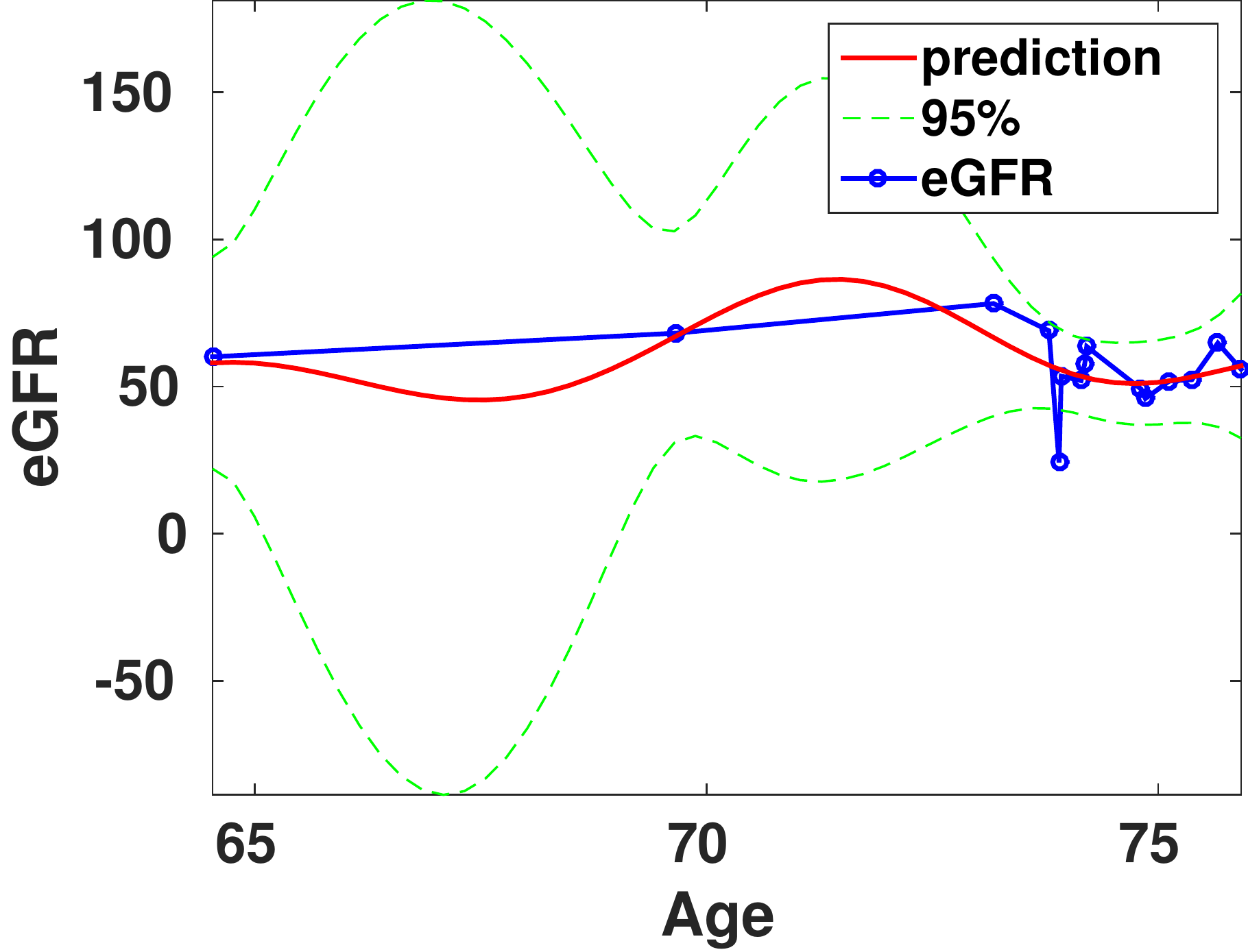}\\
\end{tabular}
\caption{The GPR model shows relatively low variance when the gap between measurements is small, but the variance increases markedly when the measurements are sparse.}
\label{varianceGPR}
\end{figure}

More flexible models such as Gaussian process regression (GPR)~\cite{rasmussen2010gaussian}, multivariate adaptive regression splines~\cite{friedman1991multivariate} and multivariate additive models~\cite{yee1996vector} can be used instead to provide the desired flexibility. For example, through the use of a kernel function GPR can avoid making the assumptions of linear regression. However, when there are gaps between the data, as is often the case with biomedical time series, the estimated variance of the predicted output can `explode'~\cite{tirunagari2016automatic} (Figure~\ref{varianceGPR}). However, these models are less interpretable, and therefore lose out in situations where a clinician simply needs to know whether a patient's condition is progressing or improving.

In order to strike a balance between interpretability and flexibility, broken-stick regression, also known as segmented or piece-wise regression, can be used to linearly model local trends~\cite{faddy2000follicle,fattorini2005simple,ritzema1994drainage,toms2003piecewise}. However, in this formulation local discontinuities are introduced at the segment boundaries, resulting in a loss of smoothness and consequently in the ability to infer trends in the boundary regions. To address this we take a Bayesian approach to derive a long-term trend by enforcing a smooth transition between the locally linear line segments, while still preserving the local trends. The ability to capture both long- and short-term trends makes this approach ideally suited to modelling biomedical time series in a clinical context. Additionally, by enforcing smoothness local rates of change can be derived, giving clinicians an indication of whether a patient's condition is progressing or not. Finally, a broken-stick model can accommodate gaps in a time series through choosing the length of each line segment in a manner that ensures that there are a sufficient number of measurements within each segment and can mitigate overfitting as it fits only locally linear line segments.

\section{Methodology}
\label{methods}
Here, $\mathbf{X}$ is used to denote a vector and $\mathbf{X}[t]$ to denote the element in the vector indexed by $t$. The remainder of the notation used is given in Table~\ref{table:symbol}.

\begin{table}[h]
\centering
\begin{tabular}{r p{.3\linewidth}{l} p{.4\linewidth}{l} }
\hline
\textbf{Variable} & Domain & Meaning \\
\hline
$\mathbf{T}$ & vector of real numbers & the time domain\\
$t$ & integer & enumerator of the time domain, from 1 to $T$\\
$w$ & integer & enumerator of the window, from 1 to $W$\\
$\mathbf{U}$ & vector of integers & indices storing the beginning of window \\
$\mathbf{L}$ & vector of integers & indices storing the end of a window\\
$\mathbf{\theta}$ & model parameters\\
$\mathbf{\theta}_w$ & vector of line segment parameters\\
$\Delta_d$ & integer & window interval of length $d$ \\
$W$ & integer & number of windows \\
$\omega_1^{(w)}$ & integer & line segment gradient\\
$\omega_0^{(w)}$ & integer & line segment intercept \\
 $\mu_t^{(w)}$ & integer & mean value of the time window\\
\hline
\end{tabular}
\caption{\label{table:symbol} Notation}
\end{table}

\subsection {Windowing}
The first step in fitting the broken-stick model is the division of a time series into a number of windows. Here, windows of equal length $d$ were used across all time series, although there is no constraint requiring the windows to be of equal length across or within individual time series. The window length was determined from the data based on the intervals between measurements, as there should be at least three measurements within each window in order to avoid overfitting line segments. In general, having more measurements within each window is preferable. However, it is only possible to influence the number of measurements within a window by increasing $d$, as the number of measurements in each time series is fixed. Given that larger values of $d$ may result in local fluctuations going undetected, while smaller values of $d$ may lead to measurement noise dominating the model, the window length must be optimised for each application.

\subsection{Local Fitting}
\begin{figure}[h]
\centering\includegraphics[width=\linewidth]{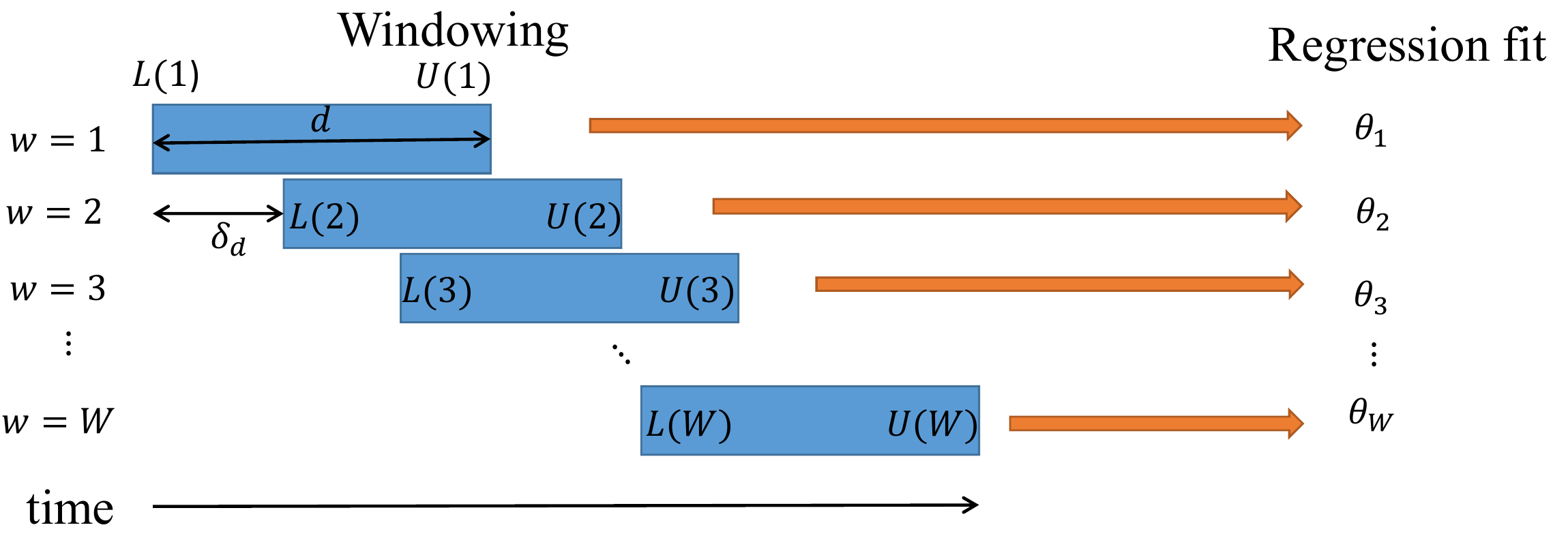}
\caption{\label{fig:windowing} A time series is broken into $W$ windows of length $d$. For each window, a linear regression model is fit.}
\end{figure}

Given $d$ and a specified interval to slide the window by, $\Delta_d$, the number of windows $W$ is also determined. For each window, a linear regression is performed by:

\begin{equation}
\label{eqn:mu_w_t}
\mu^{(w)}(t) = \omega_1^{(w)} \times t + \omega_0^{(w)},
\end{equation}

where $\omega_1^{(w)}$ is the gradient and $\omega_0^{(w)}$ is the intercept for the $w$-th window $\mathbf{\theta}_w \equiv [\omega_0^{(w)}, \omega_1^{(w)}]$. The fitting of each window is summarised in Figure~\ref{fig:windowing}.

\subsection{Window Influence}
In order to smoothly join the fitted line segment we take a Bayesian approach. Let $P(x|t)$ be the distribution of the repeated measurement $x$ given the time $t$. The local regression models will then give $P(x|w,t)$ for $w=1, \ldots, W$. These two quantities can be seen to be related by:

\begin{equation}
\label{eqn:mu}
P(x|t) = \sum_w P(x|w,t) P(w|t),
\end{equation}

where $P(w|t)$ is the posterior probability of the $w$-th window at time $t$. Ideally the further away in time a window is from $t$ the less influence its line segment has near $t$. One way to achieve this is to use Bayes' theorem to define $P(w|t)$ such that $P(w|t) \propto p(t|w)$ and the window function $p(t|w)$ is bell-shaped, e.g. is Gaussian:

\[
p(t|w) = \mathcal{N}(t|\mu_t^{(w)}, \sigma^{(w)})
\]
where $\mu_t^{(w)}$ is the mean value, i.e. the mid-point, of the time window (see Figure~\ref{fig:windowing}):
\[
\mu_t^{(w)} = \frac {[L(w)] + [U(w)]} 2  
\]
and  $\sigma^{(w)}$ is the standard deviation. This must be a function of $d$, not the window, as the standard deviation does not vary from one window to another. Here, $\sigma^{(w)} = \sigma = \frac 1 \alpha d$ so that $\alpha$ enables us to define the decay of the function in terms of distance from the mid-point of the time window.

Having defined $p(t|w)$, we can use Bayes' theorem to obtain $P(w|t)$:
\[
P(w|t) = \frac{ p(t|w) P(w)} {\sum_{w'} p(t|w') P(w')}
\]
where the prior, $P(w)$, dictates which windows should be given more weight. When a flat prior is used, we have:
\[
P(w|t) = \frac{ p(t|w) } {\sum_{w'} p(t|w')}
\]

\begin{figure}[tb]
  \centering
  \subfloat[$p(t|w)$]
  {\includegraphics[width=0.45\linewidth]{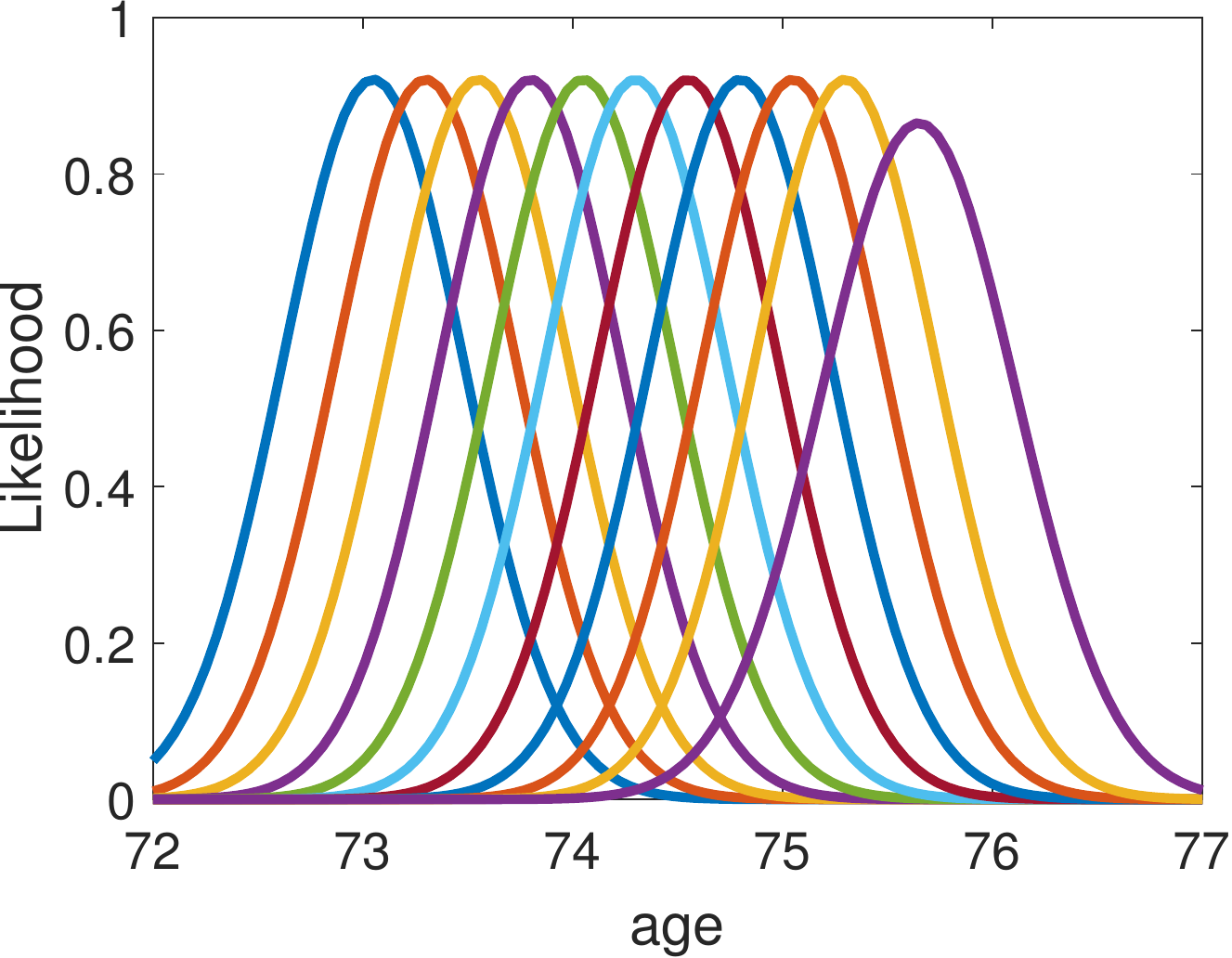}}
  \subfloat[$P(w|t)$]
  {\includegraphics[width=0.45\linewidth]{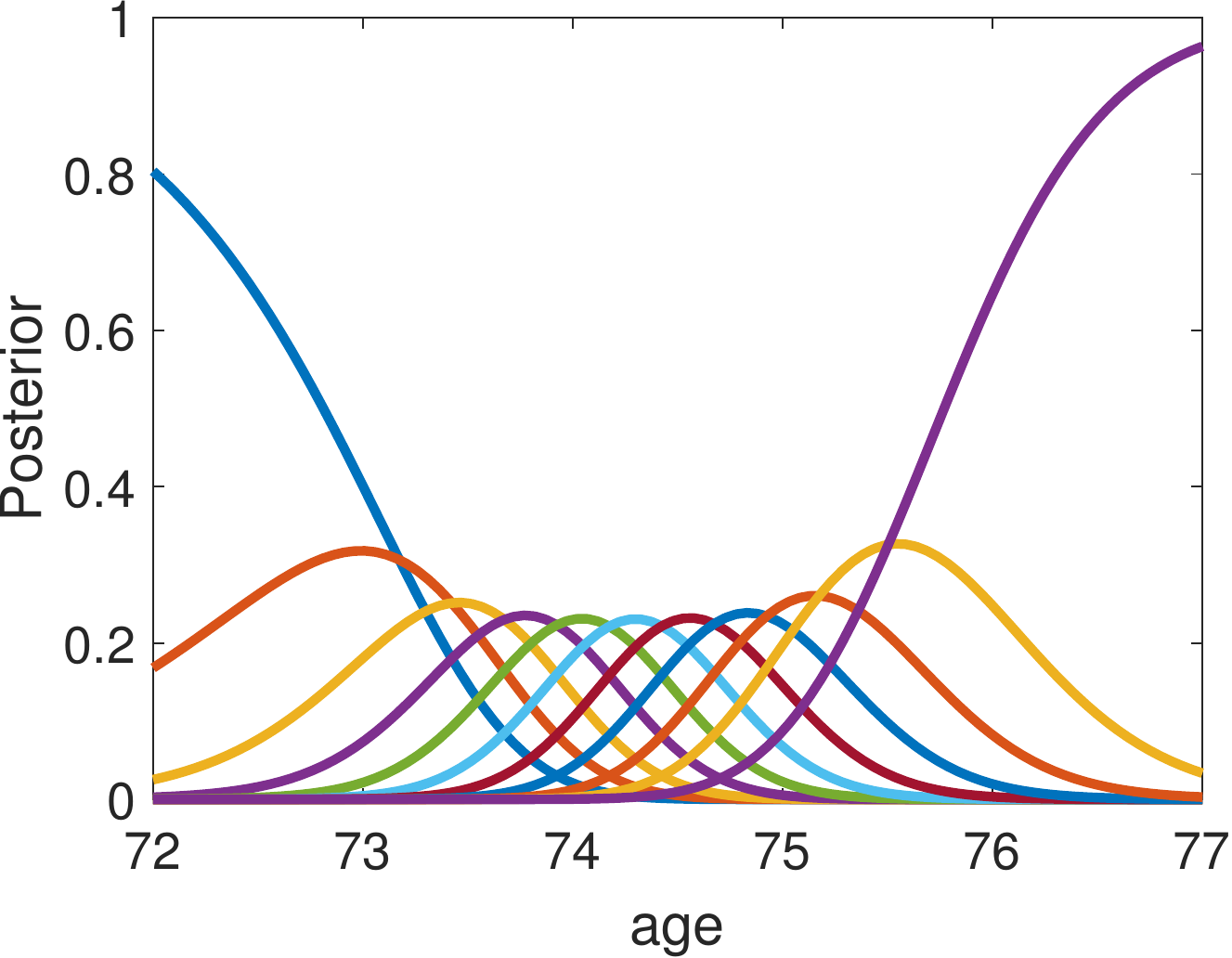}}
   \caption{\label{fig:demo_windowing} A n application of windowing with $d=2$ years and $\Delta_d = 1/2$ a year. The result for each window $w=1,\ldots, 11$ is plotted in a different colour. As the posterior probability has the property that $\sum_w P(w|t_{\ast}) = 1$, the first and last windows dominate the posterior probability near the boundaries, i.e. near $L[1]$ and $U[11]$.}
\end{figure}

The difference between $p(t|w)$ and $P(w|t)$ is depicted in Figure~\ref{fig:demo_windowing}.

\subsection{Predicted Values and Confidence Intervals}
Since the expected value of the final regression, for a given $t$, is given by 
\begin{equation}
\label{eqn:expectation}
\mu(t) \equiv E_{x\sim P(x|t)}[x] = \int x P(x|t) d_x,
\end{equation}
it follows that we can plug \Eqn{\ref{eqn:mu}} into it:
\begin{equation}
\label{eqn:mixture}
P(x|t) = \sum_w P(x|w,t) P(w|t)
\end{equation}

\begin{figure}[tb]
  \centering
  \subfloat[Local line segments]
  {\includegraphics[width=0.45\linewidth]{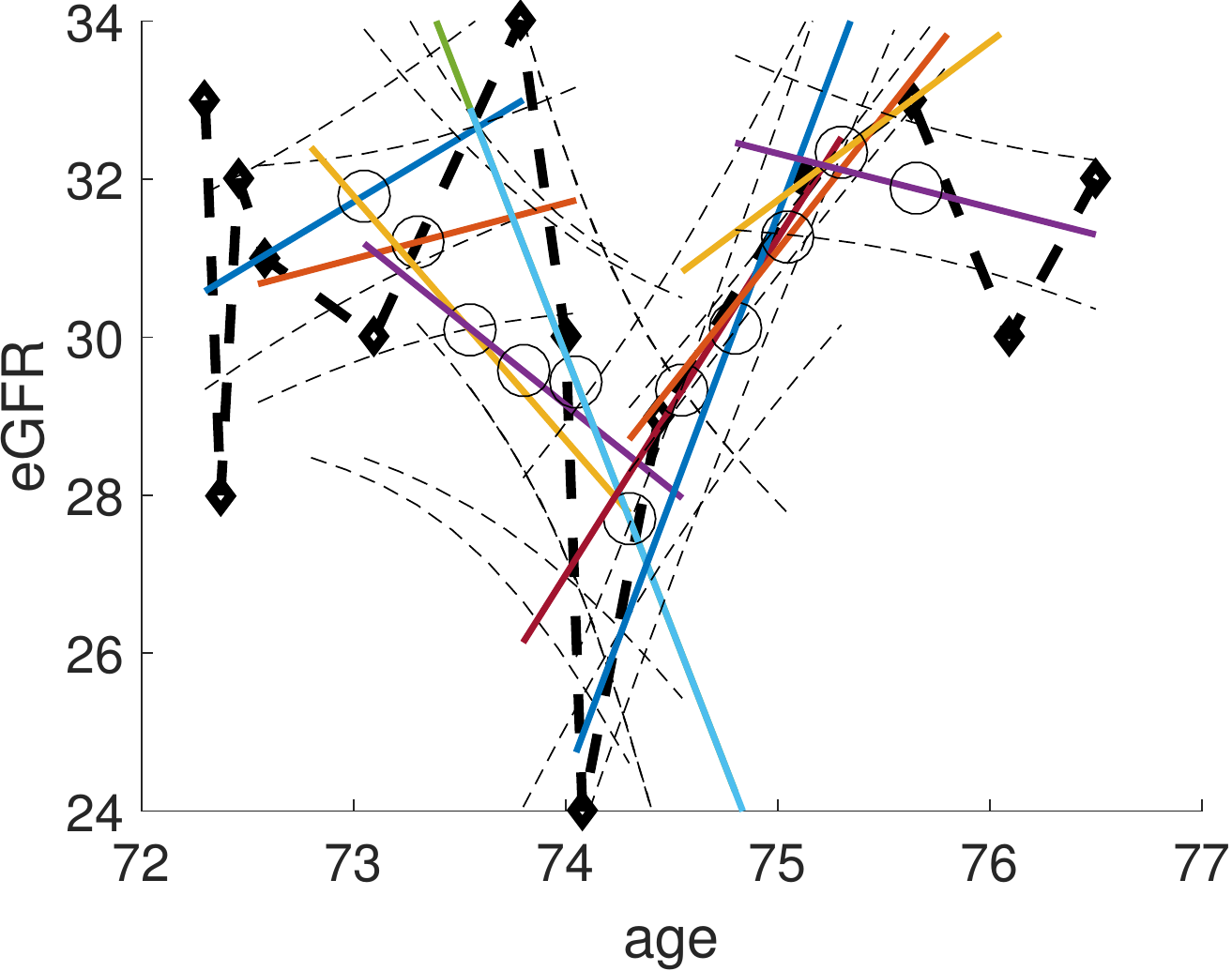}}
  \subfloat[Final model]
  {\includegraphics[width=0.45\linewidth]{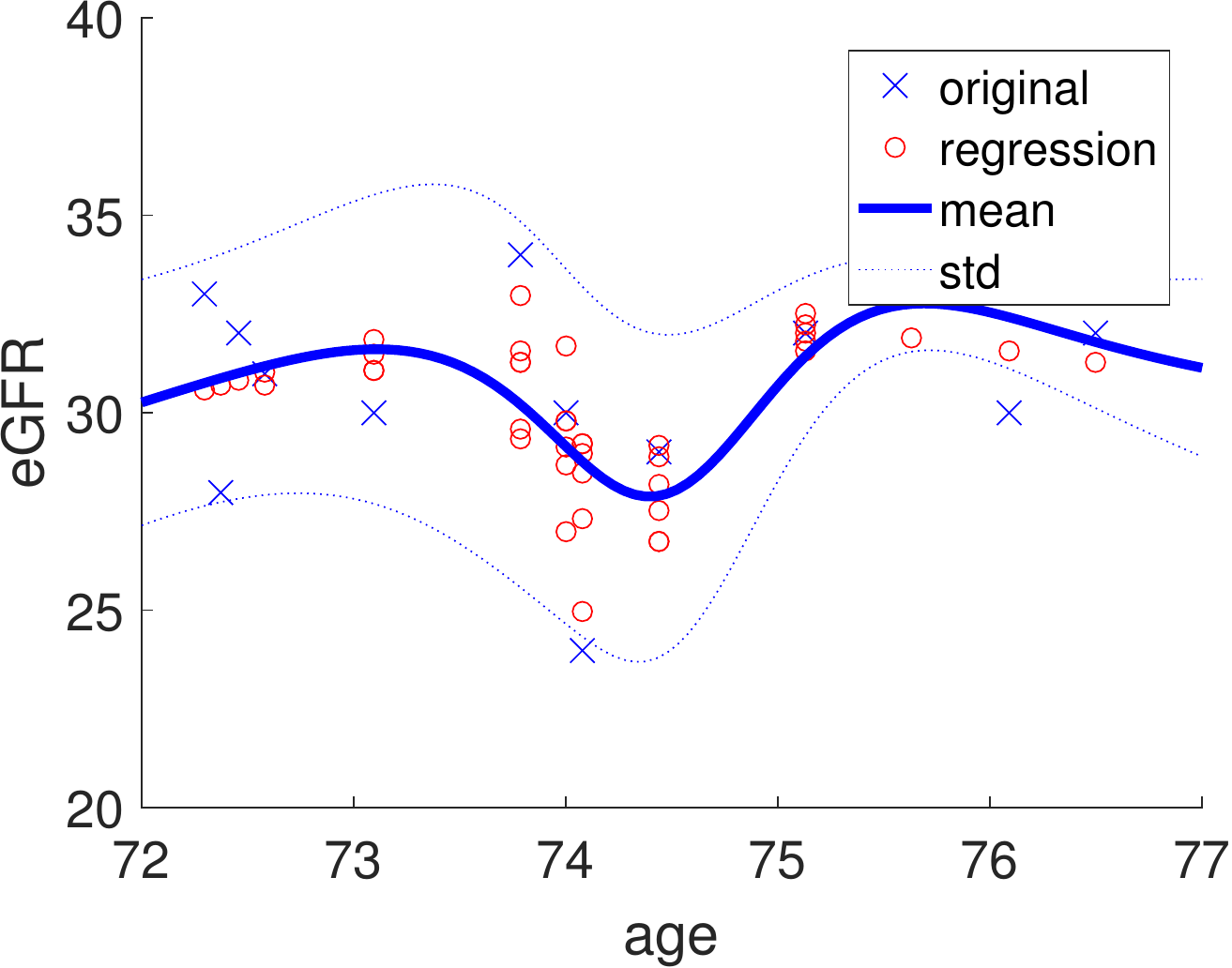}}
   \caption{\label{fig:finalModel} Fitting of a broken-stick model to an eGFR time series. (a) 11 locally linear line segments fitted to the time series. The dark dashed line represents the raw data, with the individual line segments plotted in different colours (along with their confidence intervals as dashed lines). The mid-point of each line segment is marked as an unfilled black circle. (b) The final predicted mean value $\mu(t)$ (blue line) with its confidence intervals (dashed lines). The red circles show the predicted local mean values $\mu^w(t)$ at the time points $t$ where actual eGFR measurements occur. The final fitted curve can be seen to be globally non-linear despite its locally linear constituent lines.}
\end{figure}

Taking the expectation $E_x$ on both sides of the equation we obtain:
\begin{equation}
\label{eqn:mu_t}
\mu(t) = \sum_w \mu^{(w)}_t P(w|t).
\end{equation}
Therefore, the global mean regression function is a weighted sum of the local mean regression functions, with weights determined by $P(w|t)$ at any given time point $t$. Taking approximately 95\% of the probability mass, the intervals around the global mean can be defined by:

\begin{equation} \label{eqn:mu_t_L}
\mu^{L}(t) = \sum_w \left (\mu^{(w)}_t - \sigma^{(w)}_t \right ) P(w|t)
\end{equation}
and
\begin{equation} \label{eqn:mu_t_U}
\mu^{U}(t) = \sum_w \left ( \mu^{(w)}_t + \sigma^{(w)}_t \right ) P(w|t)
\end{equation}
respectively, for the lower and upper intervals. Therefore, we have $\mu^L(t) \le \mu(t) \le \mu^U(t)$. The local line fitting for a patient's time series can be seen in Figure~\ref{fig:finalModel}(a), with the final fitted curve in (b).

\subsection{Computing the Rate Change}
\begin{figure}[H]
  \centering
  \subfloat[]
  {\includegraphics[width=0.45\linewidth]{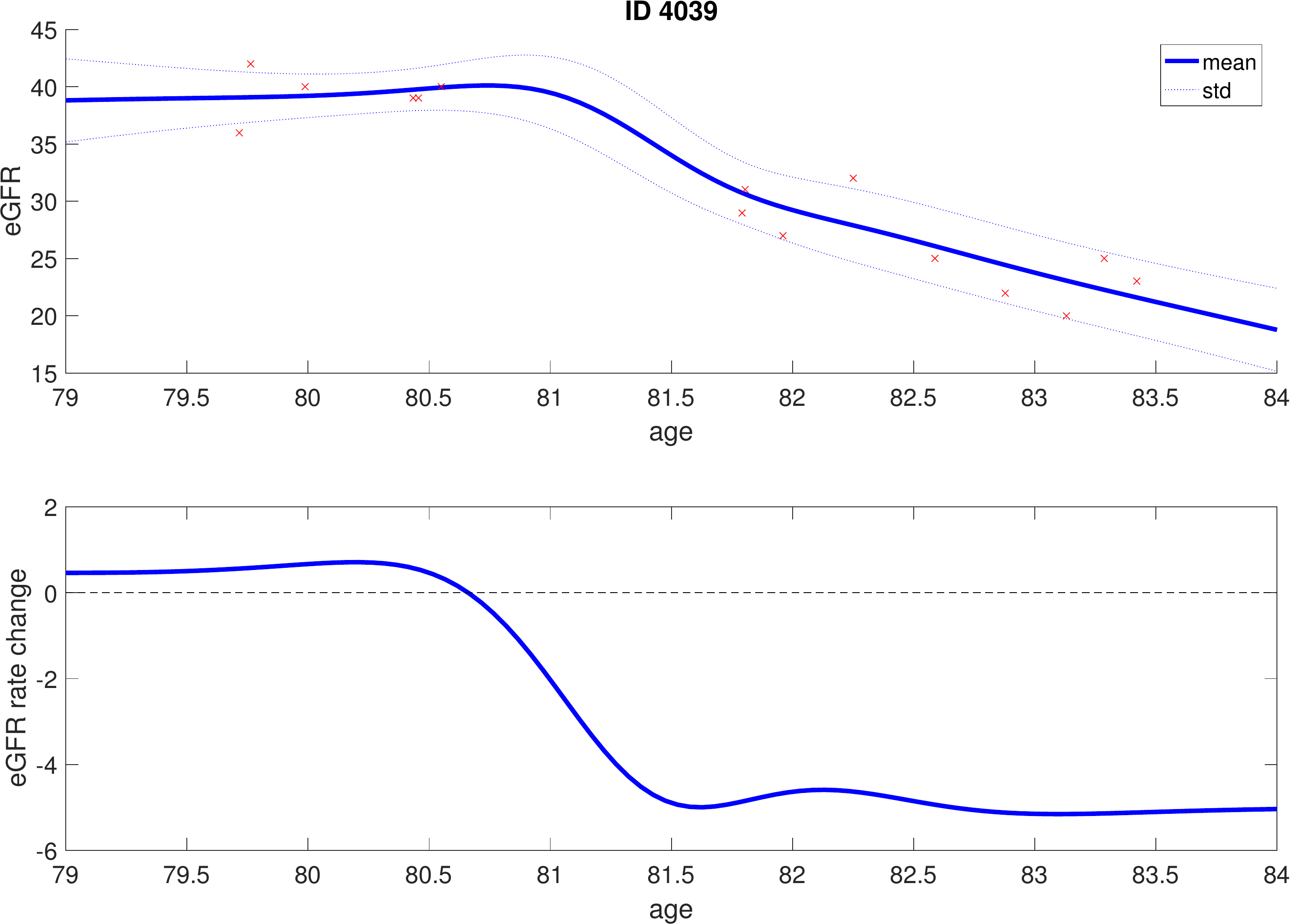}}
  \subfloat[]
  {\includegraphics[width=0.45\linewidth]{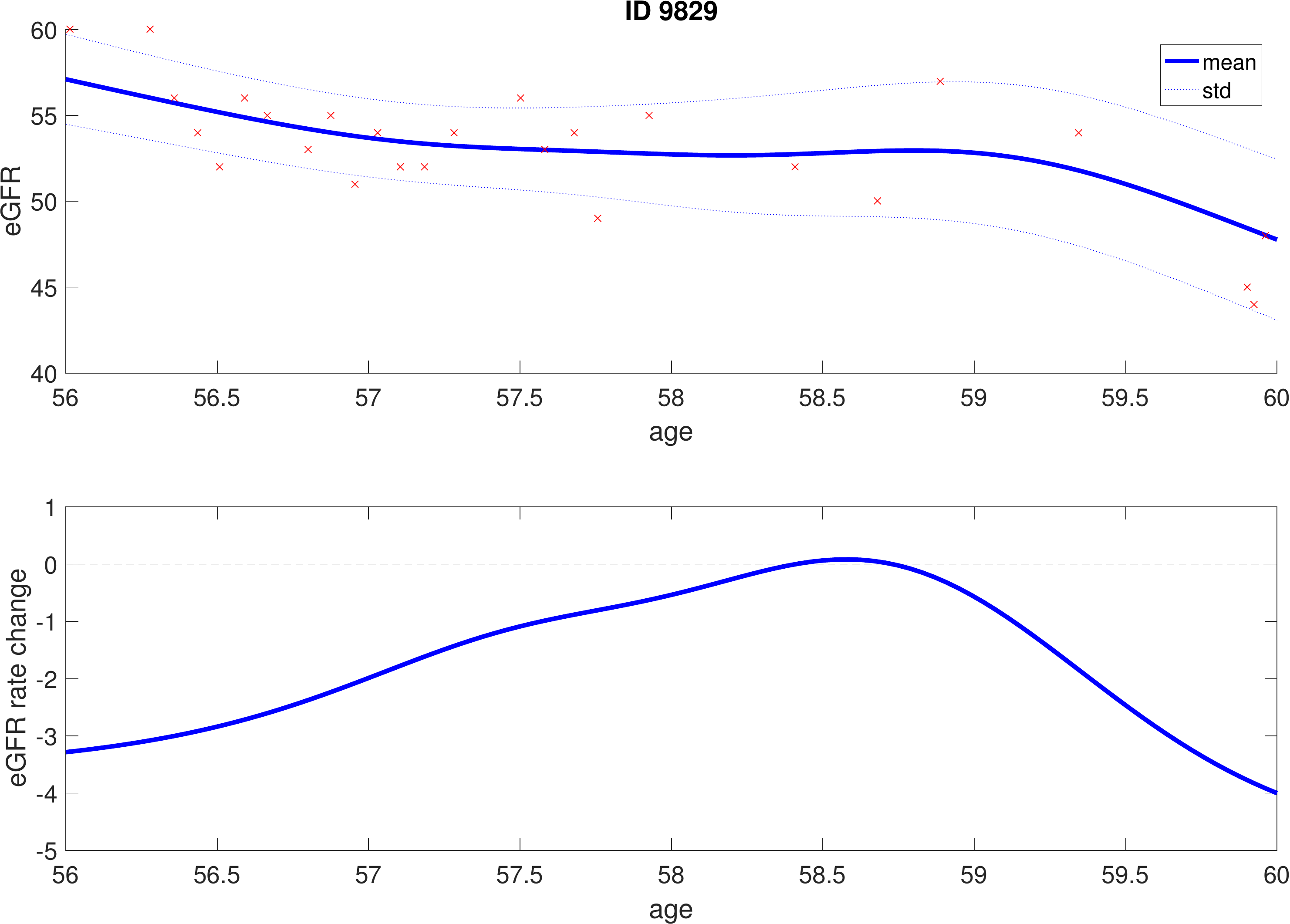}}
  
  \subfloat[]
  {\includegraphics[width=0.45\linewidth]{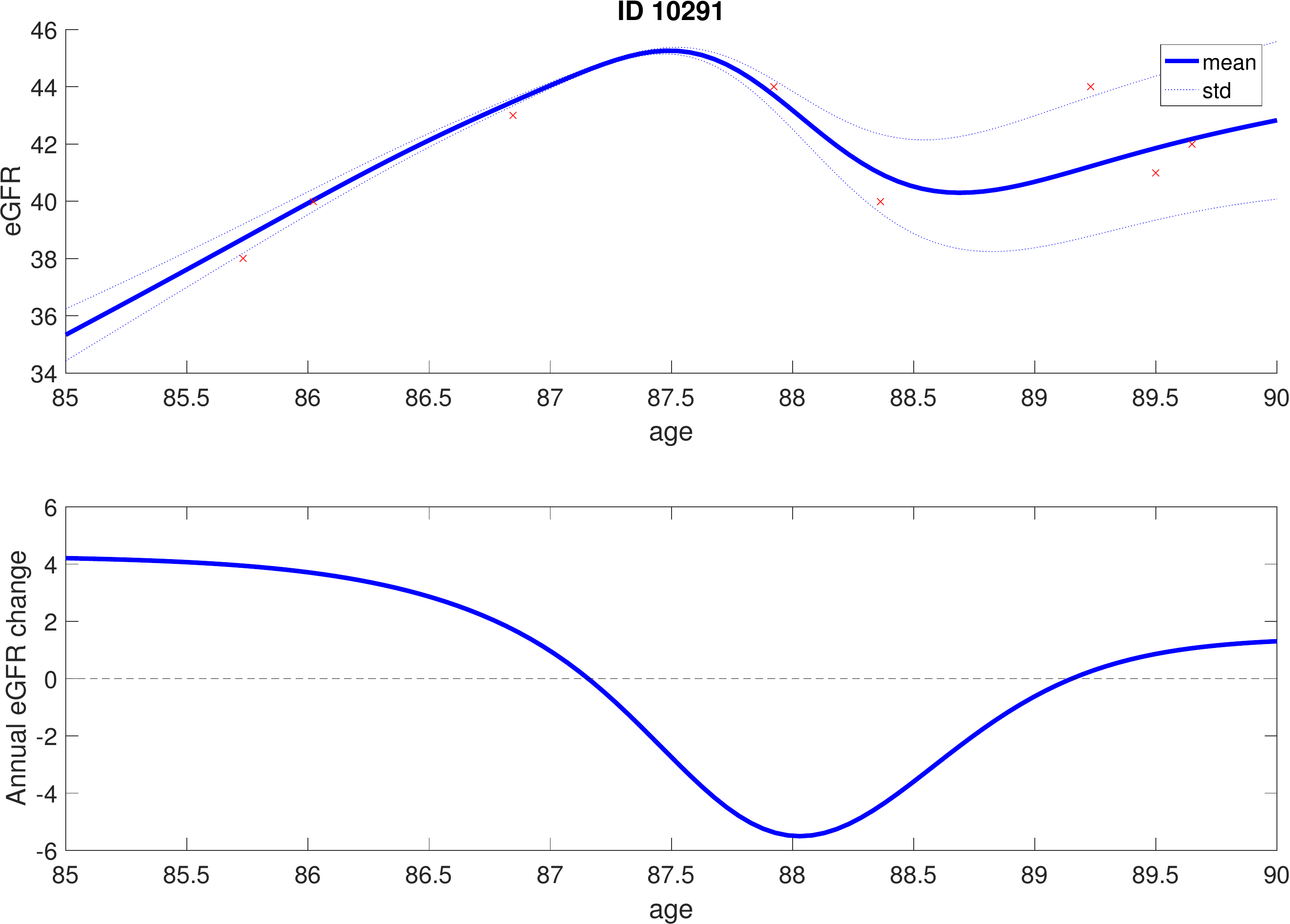}}
  \subfloat[]
  {\includegraphics[width=0.45\linewidth]{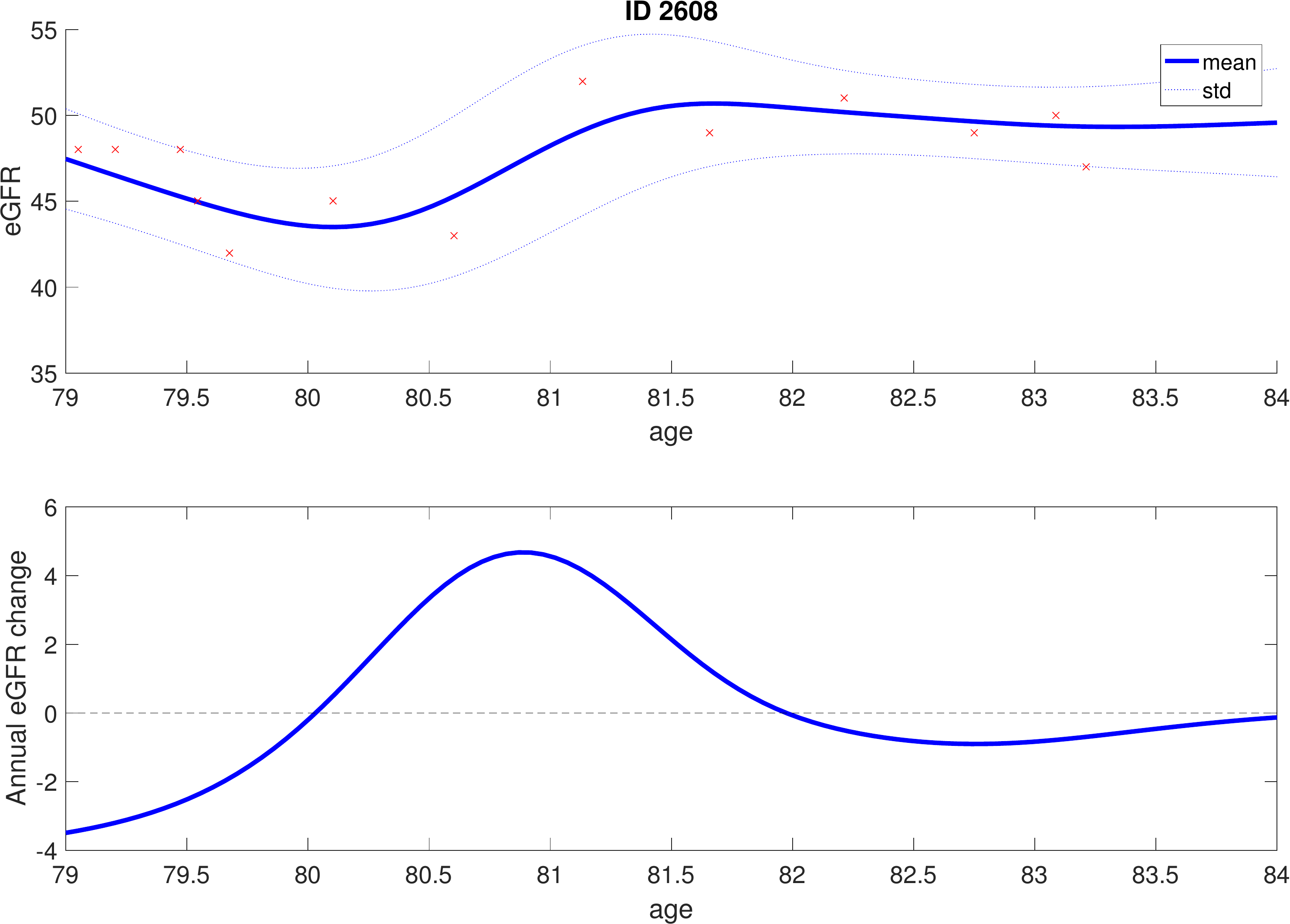}}
   \caption{\label{fig:rate_change} Four examples of eGFR time series modelled using the broken-stick model (upper diagram) along with their corresponding annual rate change, i.e. the slope of the fitted model, (lower diagram).}
\end{figure}

Another useful quantity that can be derived from the global model parametrically is the annual rate change of the time series. To do so, for each of the $W$ line segments fitted using \Eqn{\ref{eqn:mu_w_t}}, we first compute its first derivative $\omega_1^{(w)}$. The rate change can then be computed as:
\begin{equation}
\label{eqn:rate_change}
\mu'(t) = \sum_w \omega_1^{(w)} P(w|t)
\end{equation}
noting that due to it being a linear function, $\omega_1^{(w)}$ does not change with the time. The global mean trend of a patient's eGFR, given by \Eqn{\ref{eqn:mu_t}}, and the slope, computed using \Eqn{\ref{eqn:rate_change}}, can be seen in Figure~\ref{fig:rate_change} for four randomly selected patients.

\subsection {Overall Algorithm}
The overall algorithm consists of two phases, namely the model fitting phase (Algorithm~\ref{alg:training}) and the inference phase (Algorithm~\ref{alg:testing}). Following the model fitting, the calculated model parameters $\mathbf{\theta} \equiv \{\theta^{(w)}|w=1,\ldots, W\}$ and window bounds, $\mathbf{U}$ and $\mathbf{L}$, can be used in the inference phase to obtain the trend and slope for the time series.

\begin{algorithm}[tb]
%\algsetup{linenosize=\tiny}
%\scriptsize
\caption{\label{alg:training} Model fitting}
    \tabcolsep=0pt
    \begin{tabular}{@{}ll}
        \KwInput{}  & Upper and lower window bounds, $U,L $\\
                     & Repeated measurements, $\{x_t|t =1,\ldots, T\}$ \\
        \KwOutput{} & $\{\mathbf{\theta}_w| w=1,\ldots, W\}$\\
	\end{tabular}
	
$W$ = length($U$)\\	\% Obtain the window length
\For{$w \in 1, \dots, W$} {
	$\mathbf{\theta}_w  = fit(\{x_{U[w]}, \ldots, \{x_{L[w]} \})$ \\
}
\end{algorithm}

\begin{algorithm}[H]
%\algsetup{linenosize=\tiny}
%\scriptsize
\caption{\label{alg:testing} Model inference}
    \tabcolsep=0pt
    \begin{tabular}{@{}ll}
        \KwInput{}  & Upper and lower time bounds, $t_\mathtt{U},t_\mathtt{L} $\\
                     & $\{\mathbf{\theta}_w| w=1,\ldots, W\}$ \\
                     & Upper and lower window bounds, $U,L $\\
        \KwOutput{} &  trend, $\{\mu_t \pm \sigma_t|t\} $\\
					& annual rate change, $\{{\mu'}_t|t\} $\\
	\end{tabular}
	
$W$ = length($U$) \% Obtain the window length\\

\% Create the window \\
\For{$w \in 1, \dots, W$} {
	$\mathbf{b}_w = \mathcal{N}(t| \frac { t_{L[w]} + t_{U[w]}} 2, \sigma ) \mbox{ for } t \in \mathbf{T}$ \\
}

\% Normalize the window weight (Equation \Eqn{\ref{eqn:mixture}}) \\
\For{$t \in 1, \dots, T$} {
	$b_w[t] = \frac {b_w[t] } { \sum_w b_w[t]}$\\
}

\% Get the local trends\\
$\mathcal{T}$ = sample($t_\mathtt{U},t_\mathtt{L} $,1000) \%draw 1000 equall-spaced samples\\
\For{$w \in 1, \dots, W$} {
	$[\mu^{(w)}(t), \sigma^{(w)}(t)]$ = infer($\mathbf{\theta}_w$), for $t \in \mathcal{T}$ \\
	$ \mu'_w(t)$ = Calculate first derivative $(\mathbf{\theta}_w)$, for $t \in \mathcal{T}$\\
}

\% Combine the local trends and the weights as in Equations~(\ref{eqn:mu_t}--\ref{eqn:rate_change})\\ 
\For{$t \in 1, \dots, T$} {
	$\mu[t] = \sum_w \mu_w[t] \times b_w[t]$ \% mean value\\
	$\mu^U[t] = \sum_w (\mu_w[t] + \sigma_w[t]) \times b_w[t]$ \% upper interval\\
	$\mu^L[t] = \sum_w (\mu_w[t] - \sigma_w[t]) \times b_w[t]$ \% lower interval\\
	$\mu'[t] = \sum_w \mu'_w[t] \times b_w[t]$ \% annual rate change \\
}
\end{algorithm}

\section{Effect of the Window Length and Interval}
\begin{figure*}
\setlength{\tabcolsep}{1pt}
\centering
\begin{tabular}{cccc}
\includegraphics[scale=0.22]{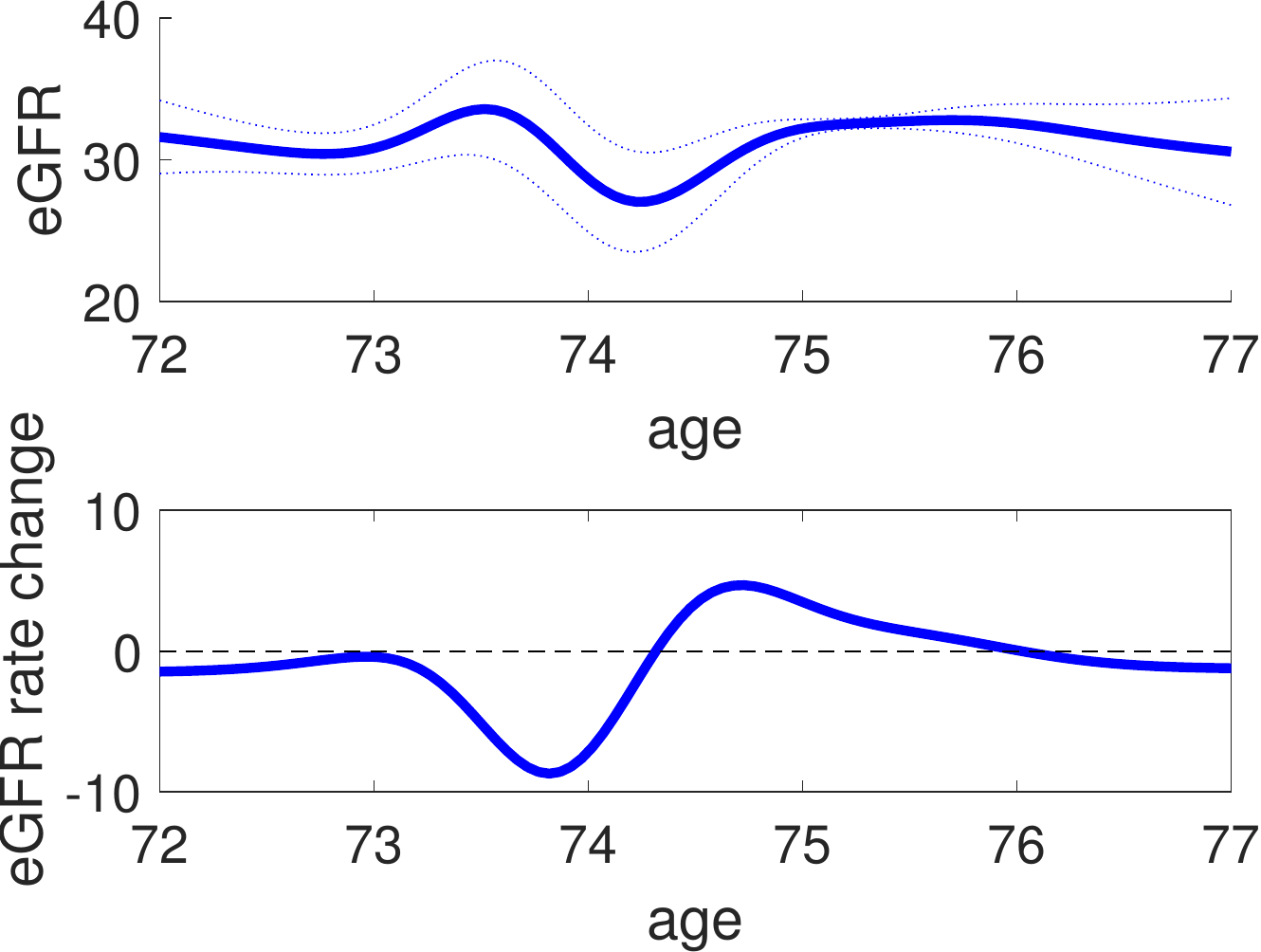} &
\includegraphics[scale=0.22]{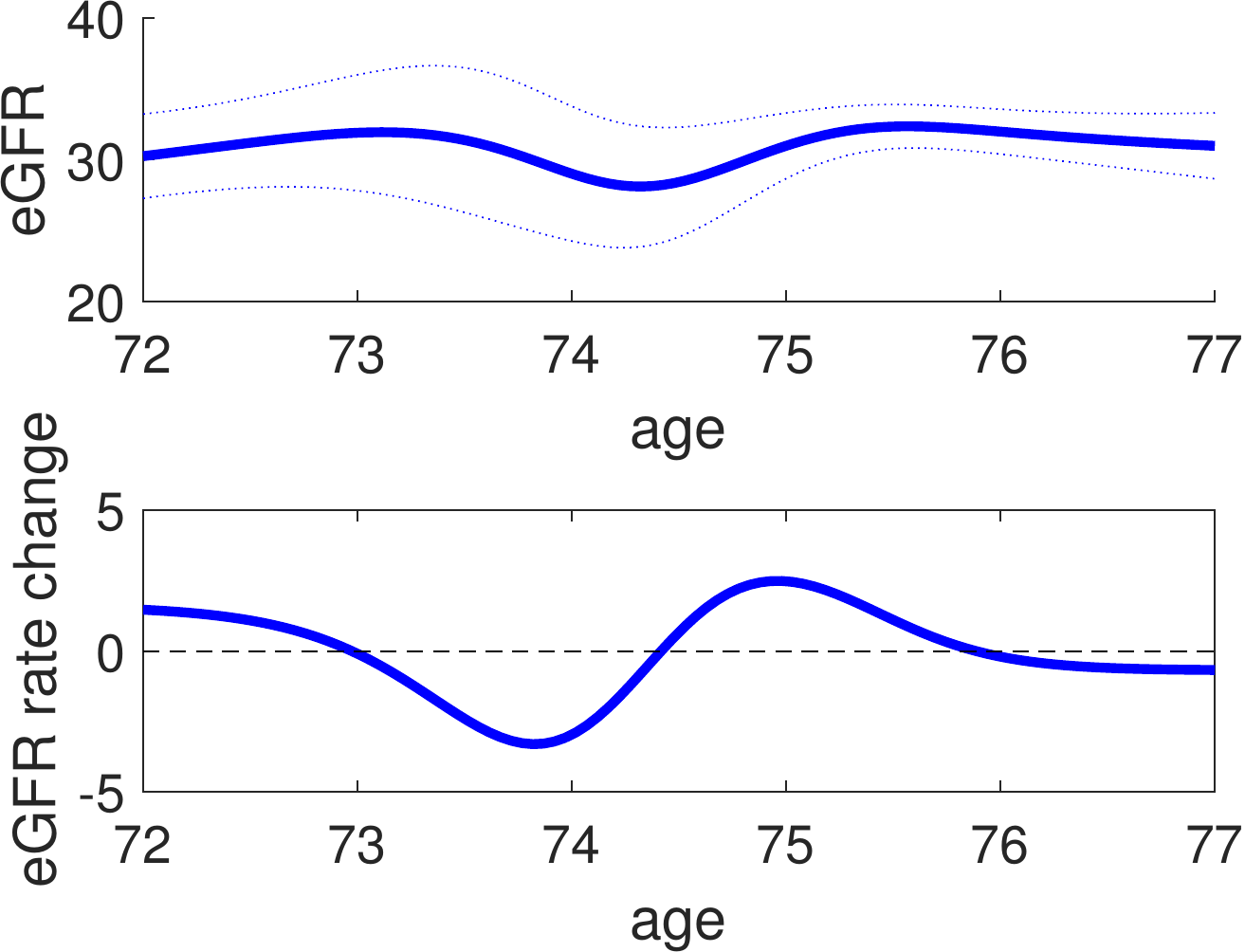} &
\includegraphics[scale=0.22]{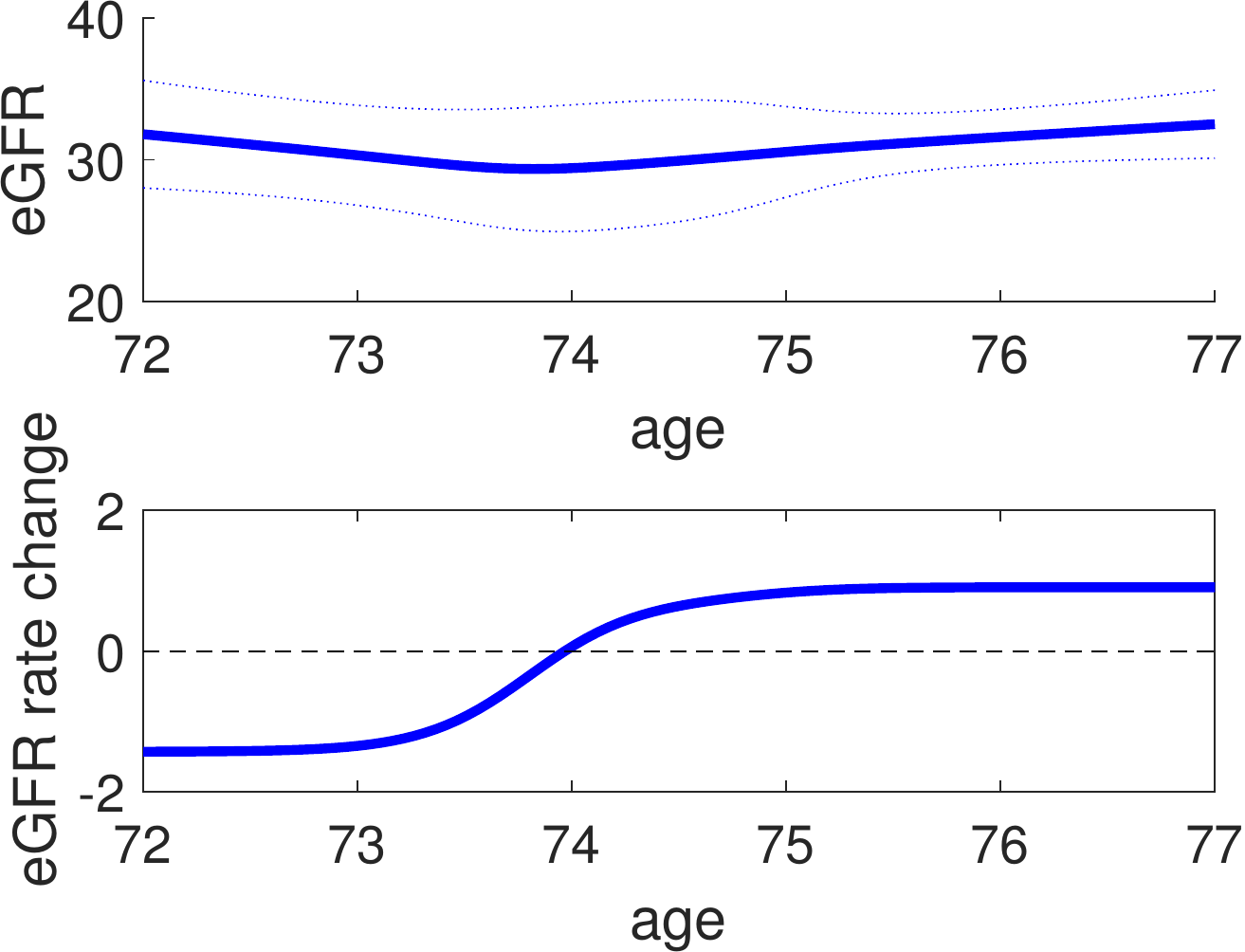} &
\includegraphics[scale=0.22]{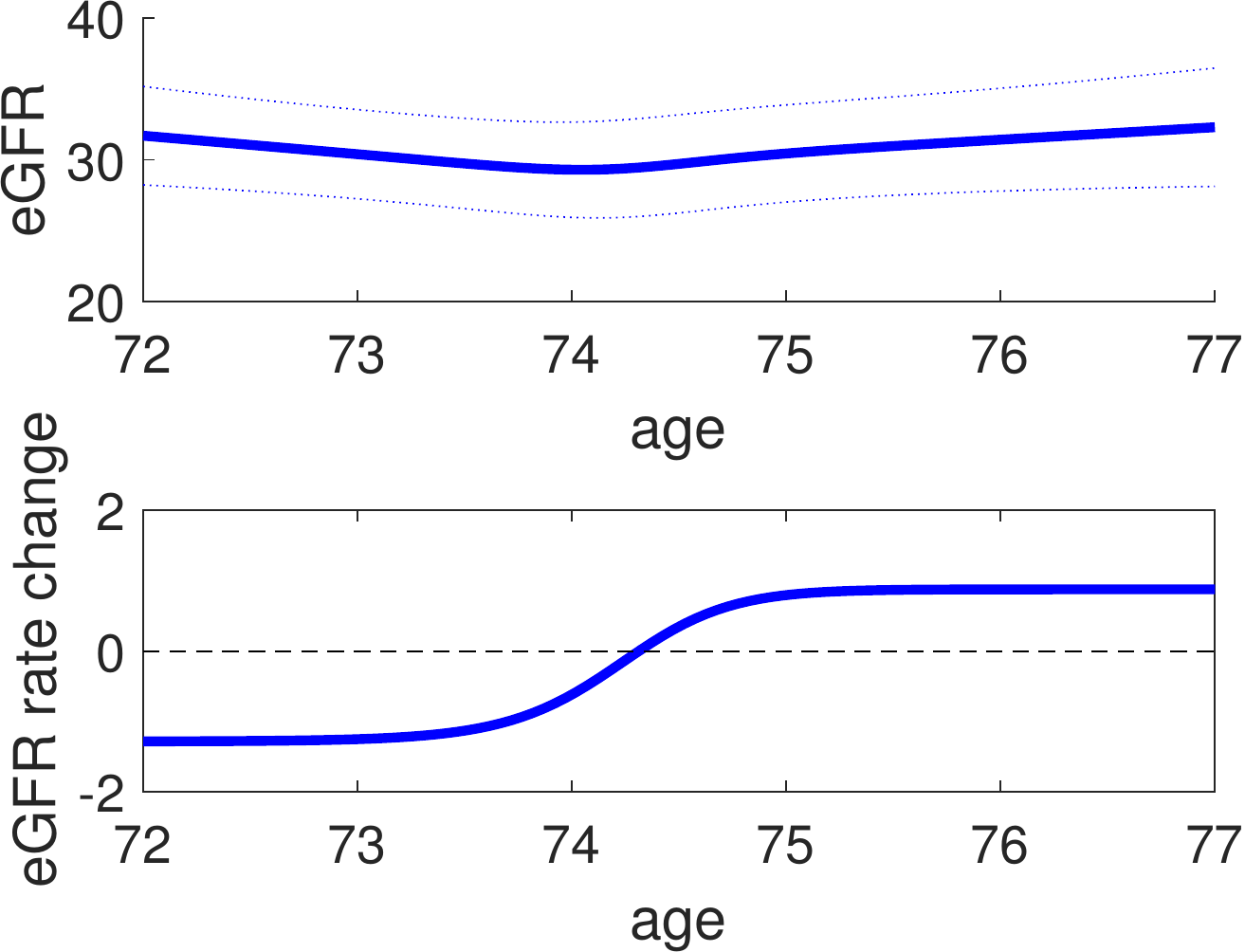} \\
\includegraphics[scale=0.22]{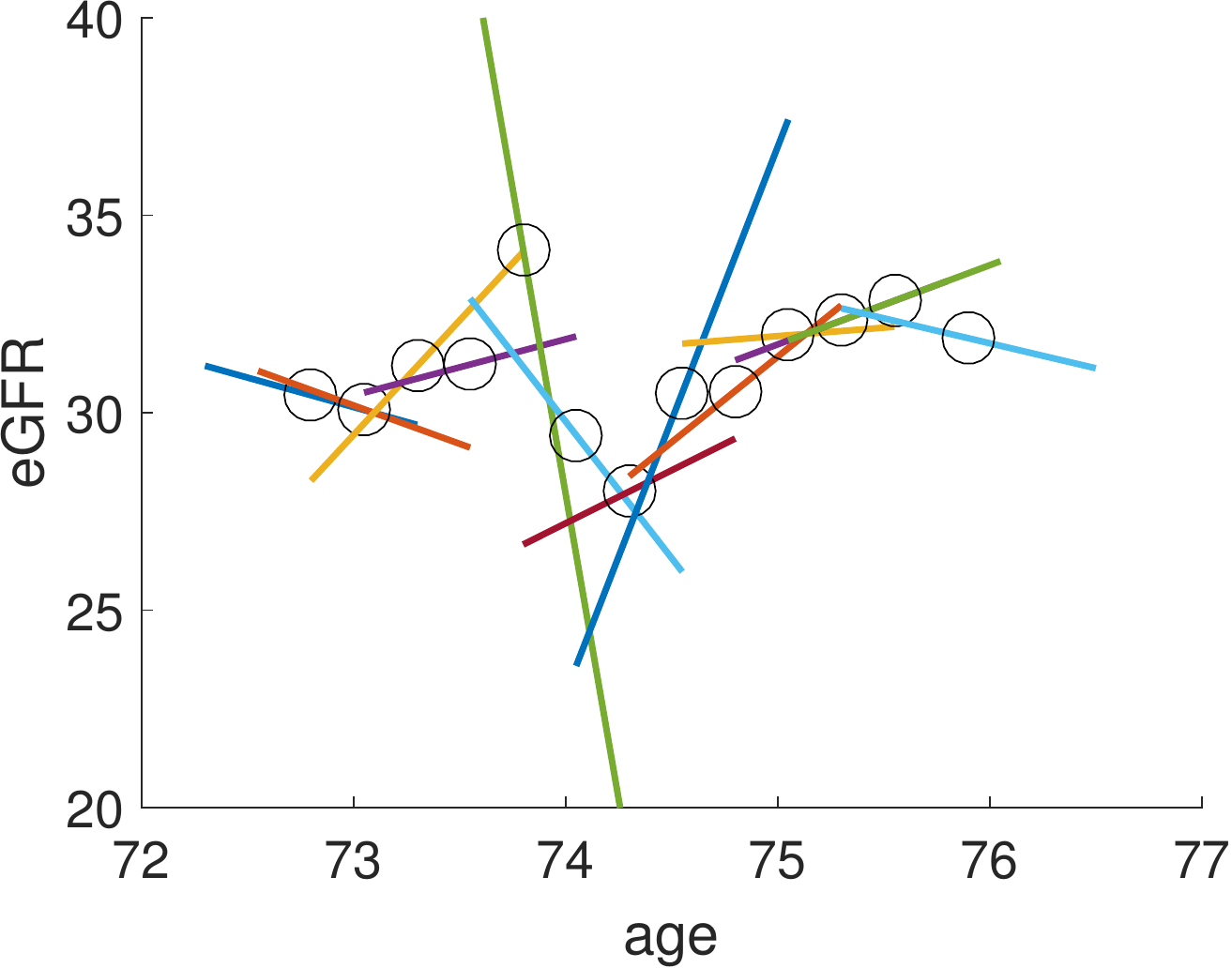} &
\includegraphics[scale=0.22]{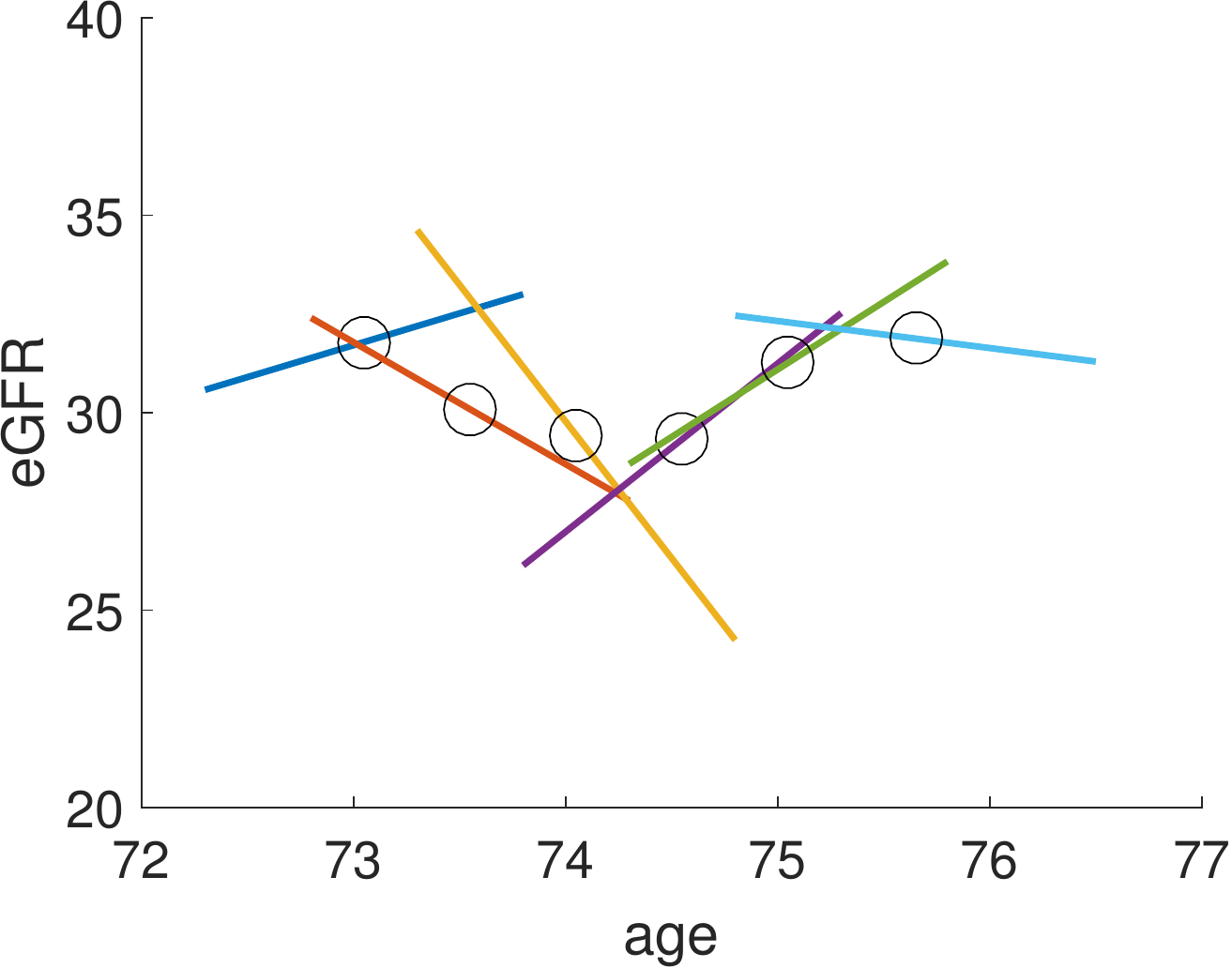} &
\includegraphics[scale=0.22]{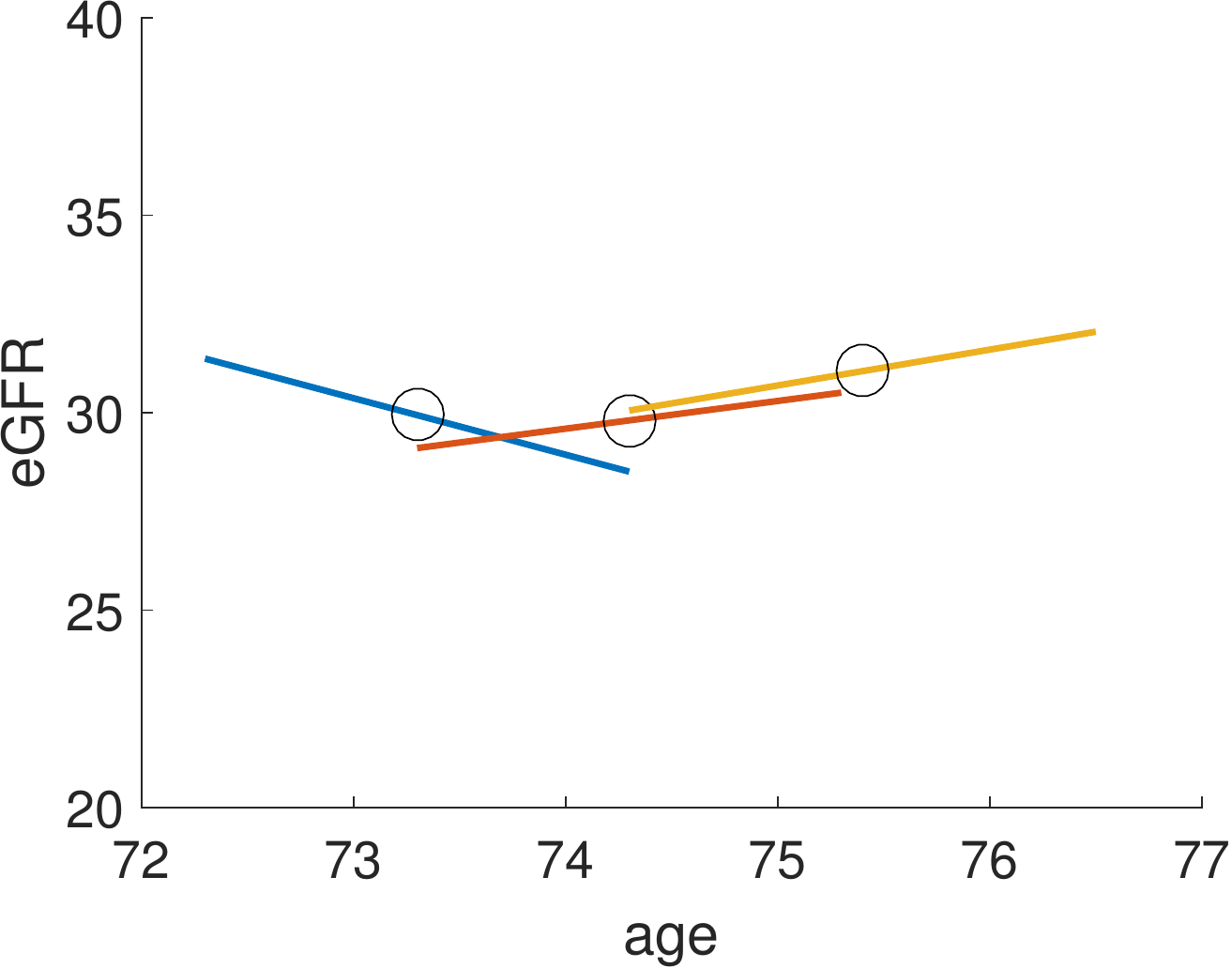} &
\includegraphics[scale=0.22]{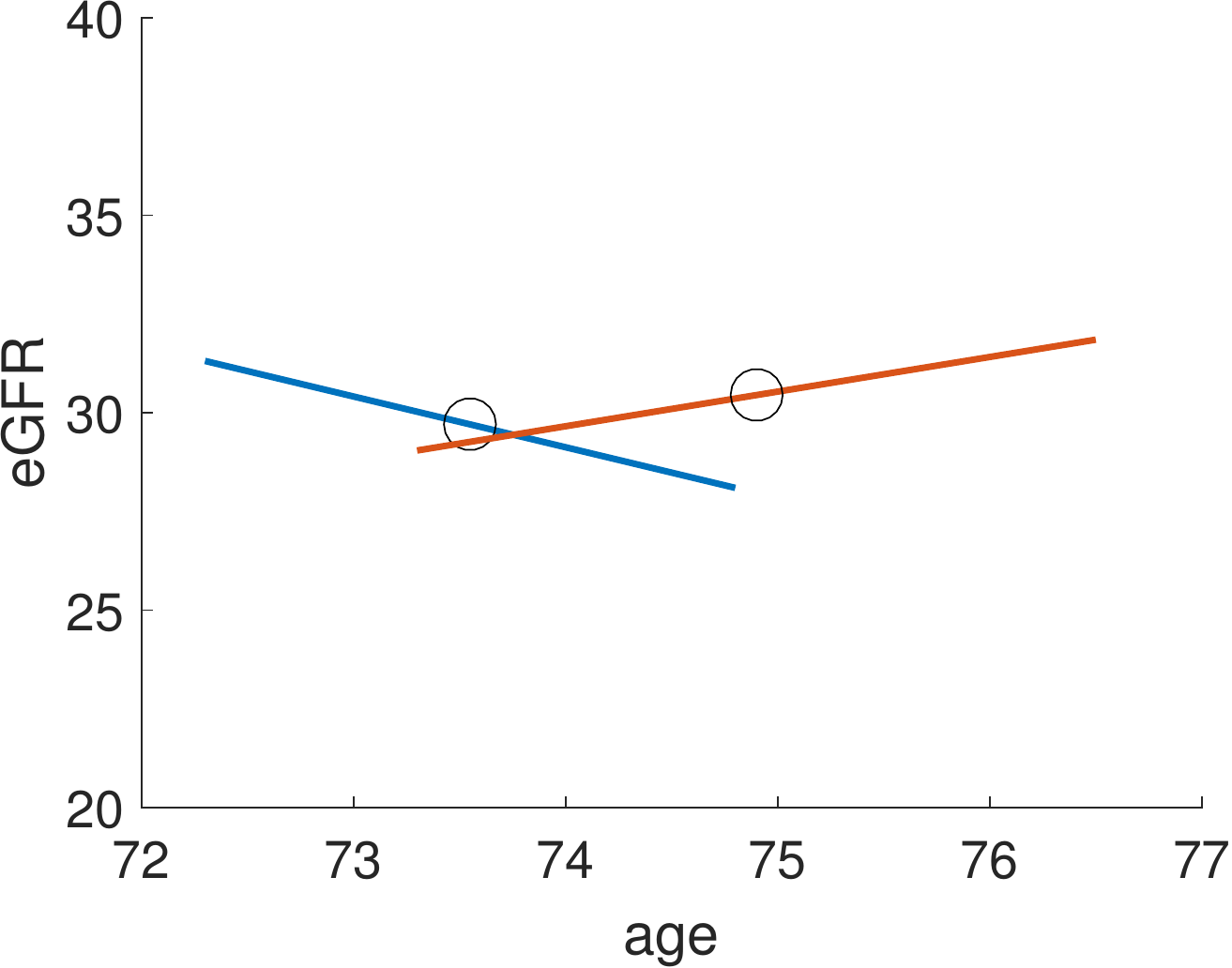} \\
(a) $(1,0.25)$ & (b) $(1.5, 0.5)$ & (c) $(2,1)$ & (d) $(2.5, 1)$\\
\end{tabular}
\begin{tabular}{ccc}
\\
\includegraphics[scale=0.22]{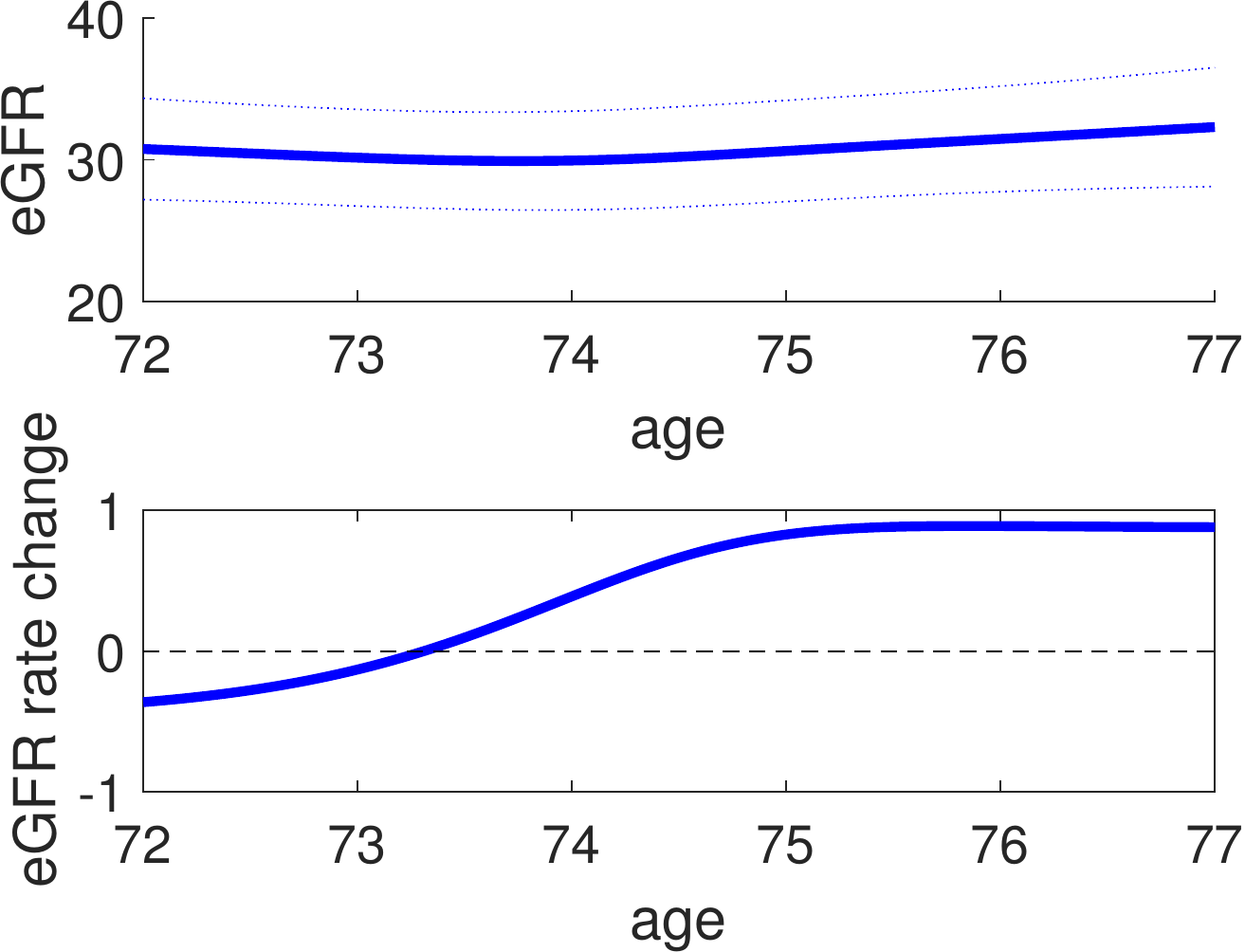} &
\includegraphics[scale=0.22]{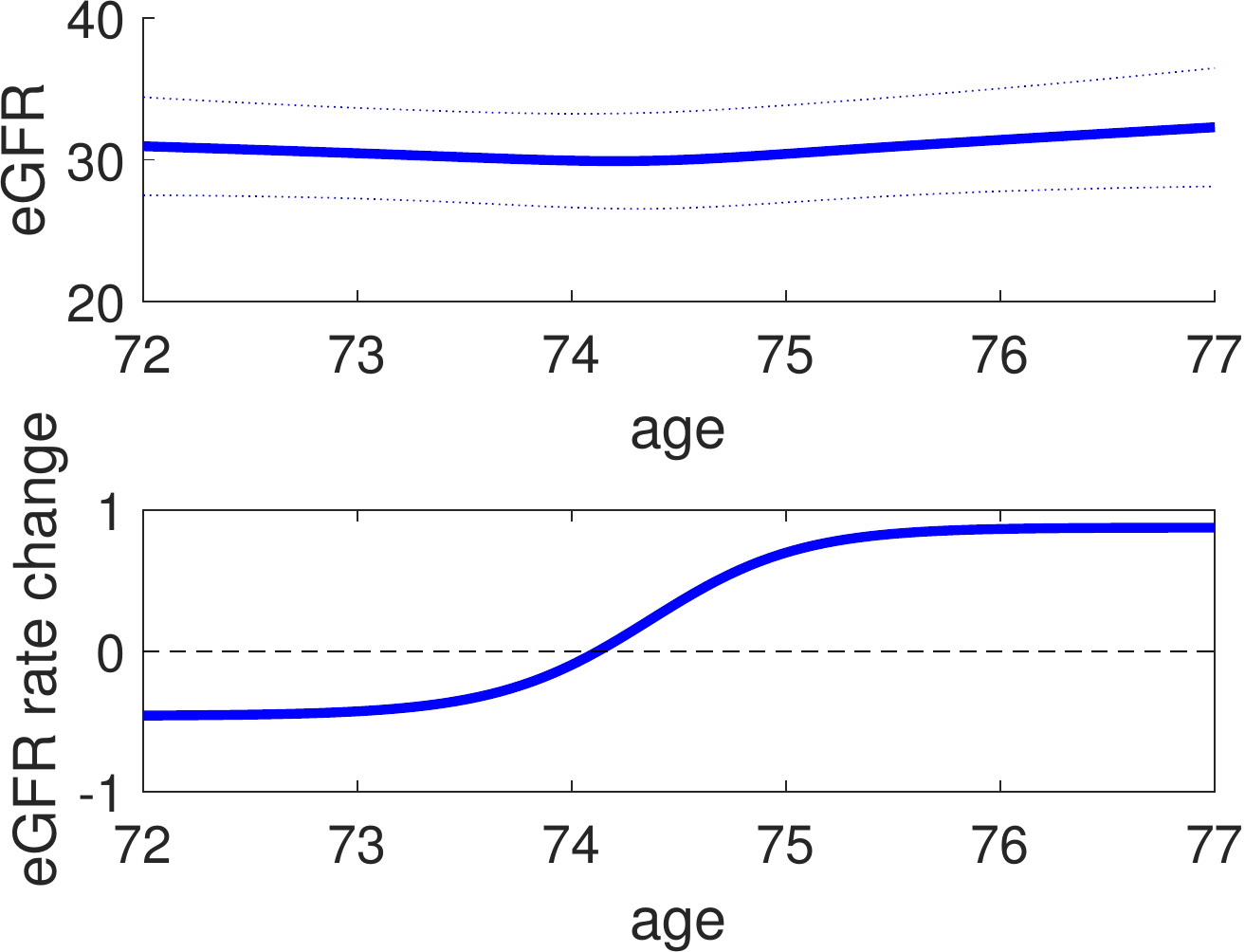} &
\includegraphics[scale=0.22]{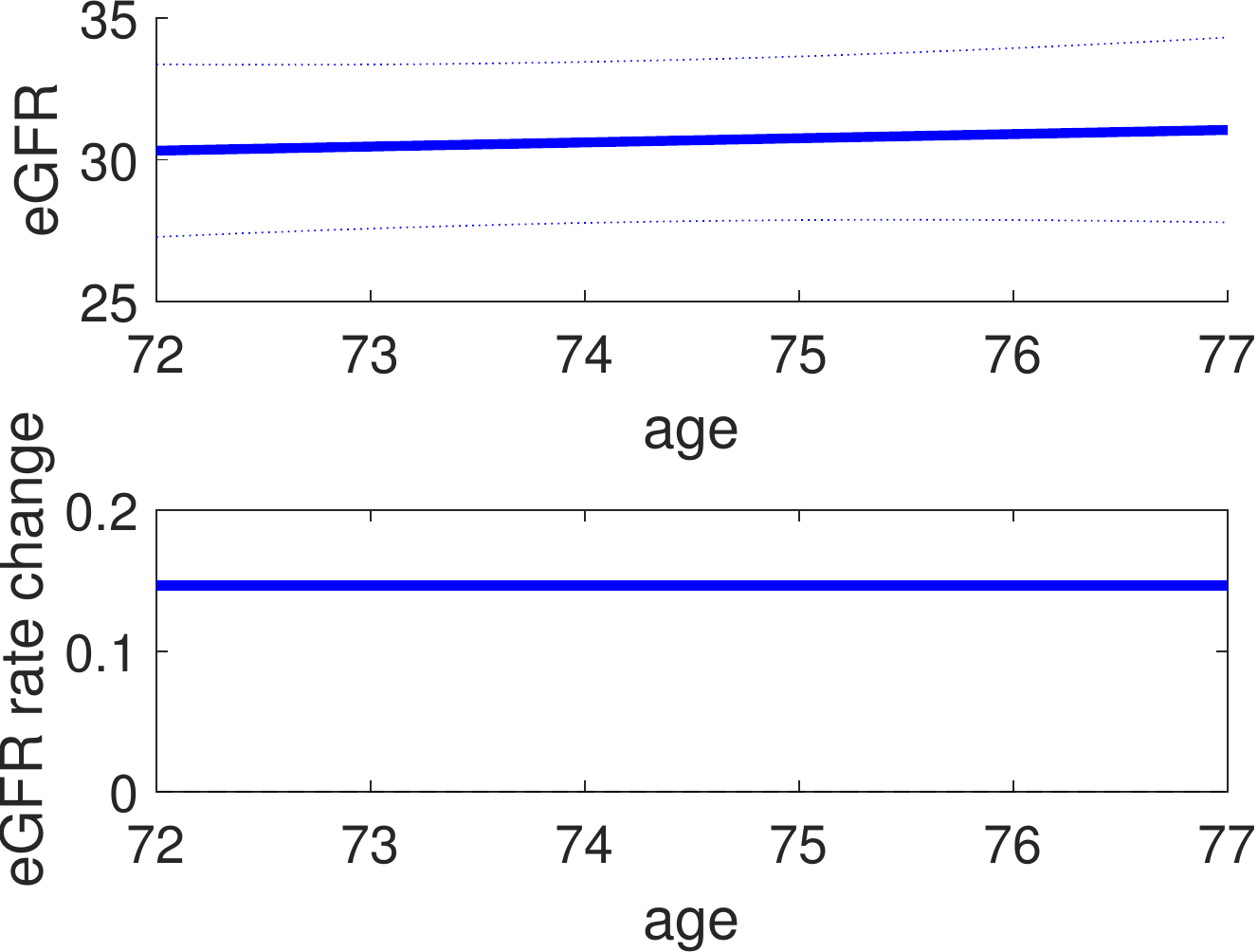} \\
\includegraphics[scale=0.22]{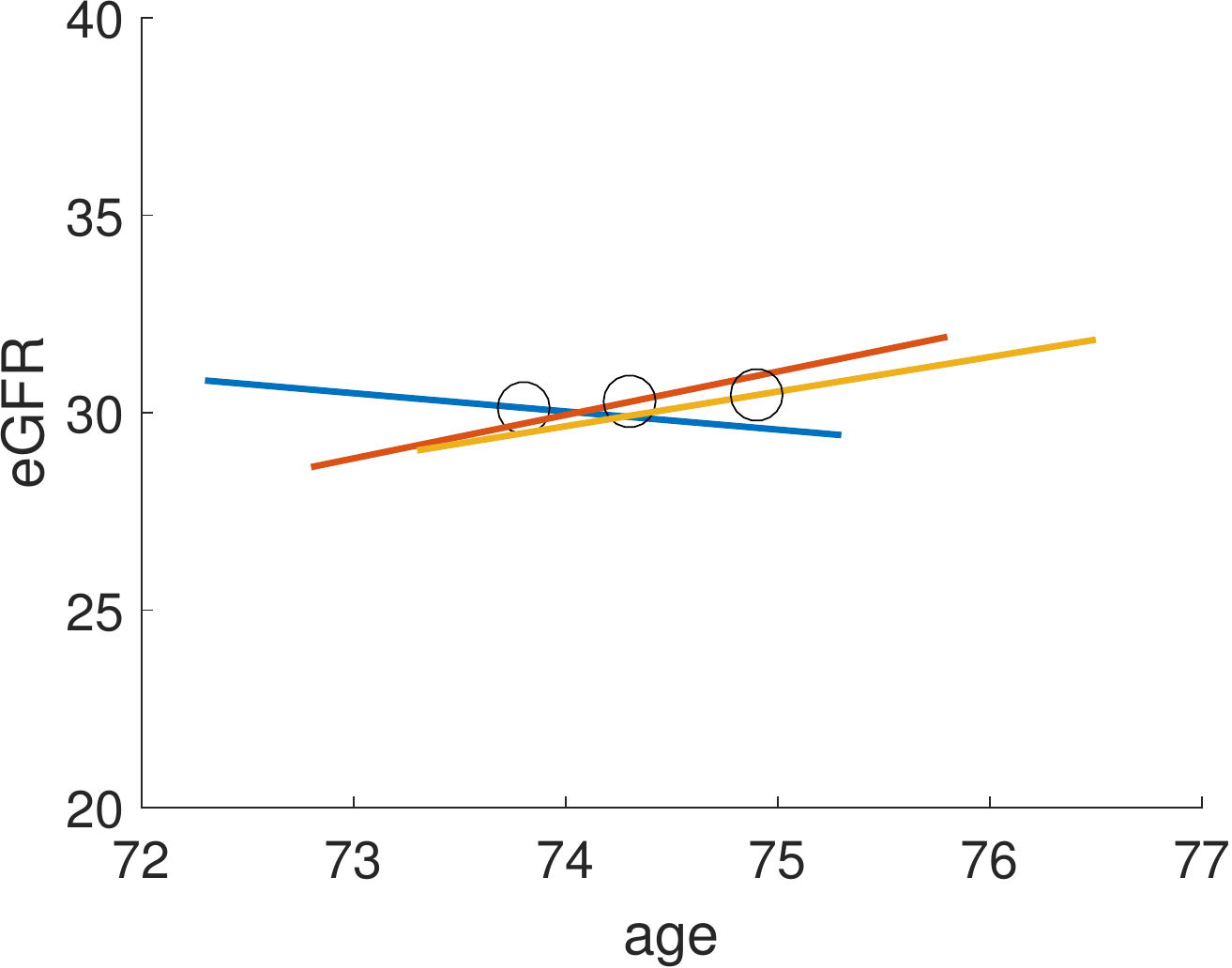} &
\includegraphics[scale=0.22]{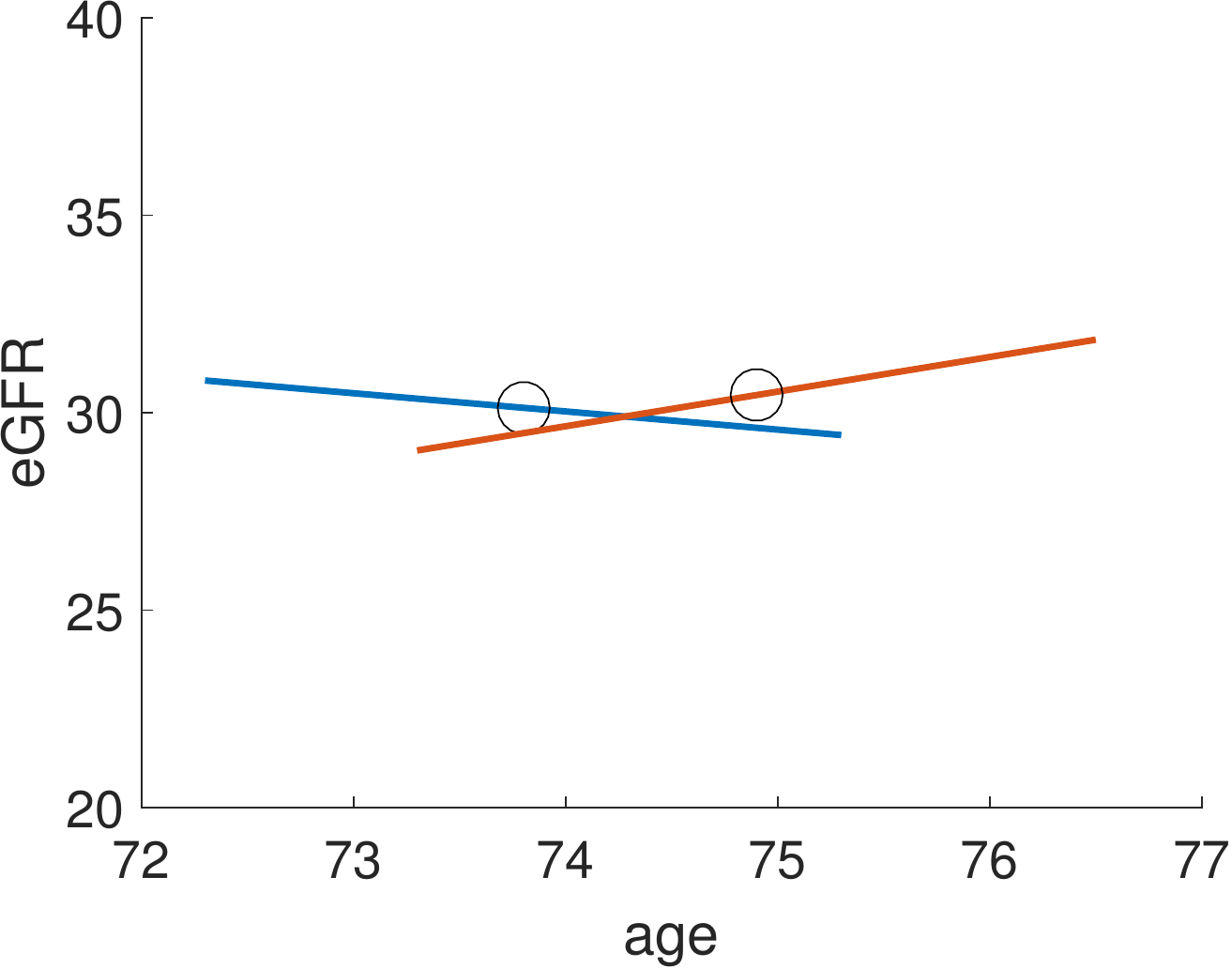} &
\includegraphics[scale=0.22]{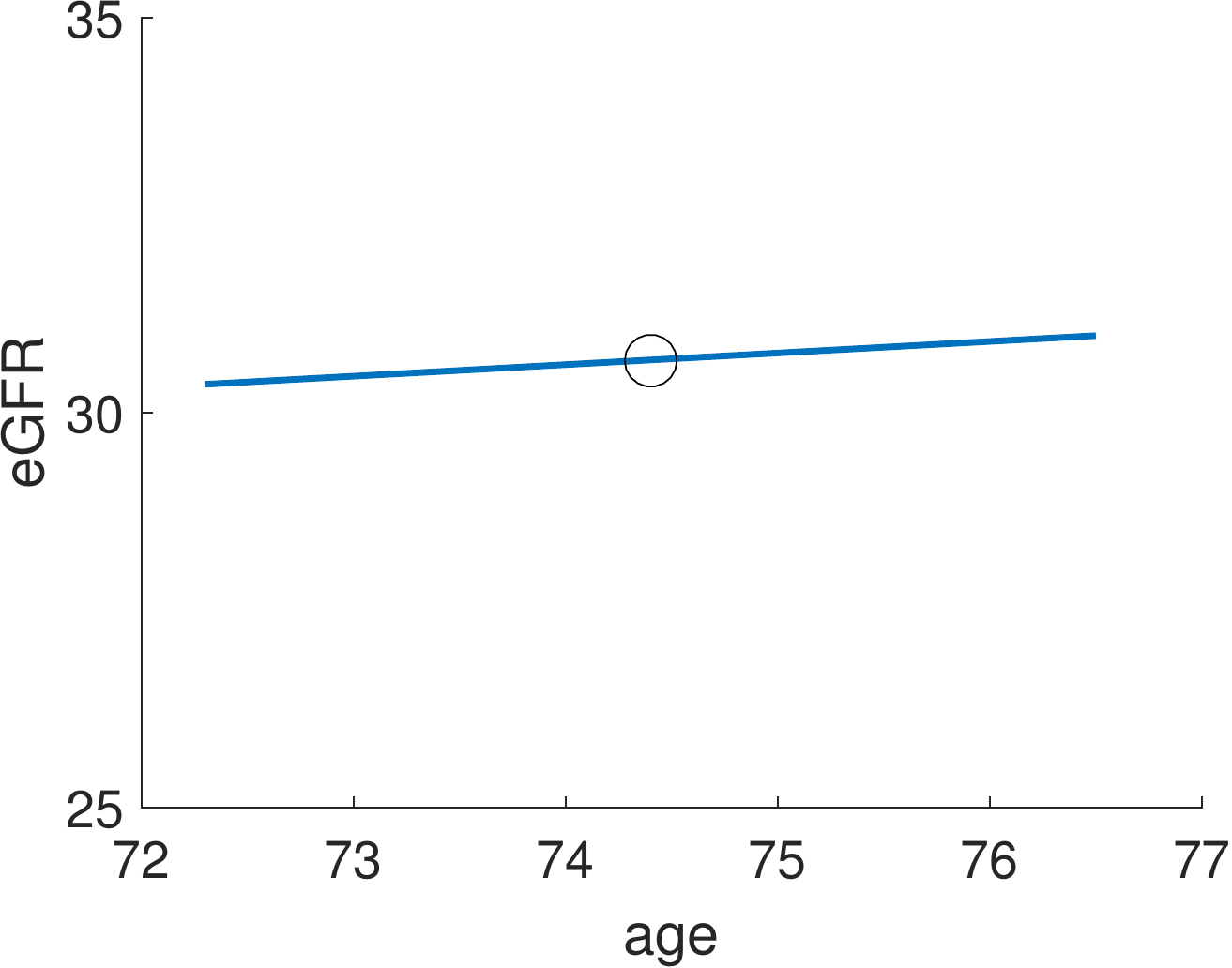} \\
(e) $(3,0.5)$& (f) $(3,1)$ & (g) $(3,2)$  \\ 
\end{tabular}
\caption{\label{fig:demo1} An illustration of fitting an eGFR time series with different window length $d$ and window interval $\Delta_d$ parameters (in years). The raw eGFR time series can be found in Figure~\ref{fig:finalModel}(a). For each figure (a)-(g) the top subfigure is the mean eGFR over time, the middle is the eGFR slope over time and the bottom the line segments of the broken-stick model.}
\end{figure*}

In order to illustrate the effect of the window length $d$ and the window interval $\Delta_d$, different parameter pairs $(d, \Delta_d)$ were used to fit an eGFR time series in which no AKI was observed. The results of this can be seen in Figure~\ref{fig:demo1}. From this it can be seen that shorter window lengths and shorter intervals produce more sensitive models, as can be seen in the differences between (c) and (f) and between (e), (f) and (g) in terms of the magnitude of the slope. The impact of choosing windows lengths and intervals that are too large can also be seen in the lack of local fluctuations in (g). From this it is reasonable to conclude that the expected eGFR and slope are only comparable between patients when the same fitting parameters are used.

\section{Case Study}
In order to demonstrate the utility of the proposed broken-stick model, we applied it to primary care data collected for the QICKD study~\cite{deLusignan2009QICKD} in order to model the long-term trend of eGFR measurements. For patients with CKD, their eGFR is one of the primary outcomes used by clinicians and is used in determining the stage, and therefore severity, of a patient's CKD. While a true clinical staging of CKD will take kidney damage, as evidenced by the level of albuminuria, into account as well as eGFR, we focus only on eGFR here in order to demonstrate the utility of the proposed broken-stick model. The possible stages of CKD as a function of a patient's eGFR value can be seen in Table~\ref{table:CKD_stage}, with the caveat that patients without CKD are defined here as having stage 0 CKD.

\begin{table}[h]
\centering
\begin{tabular}{|l|r|r|}
\cline{1-3}
CKD stage & $g_{\mathtt{L}}$ & $g_{\mathtt{U}}$  \\ \hline
0         &   120    &   $\infty$     \\ \hline
1         &   90    &   120$\dagger$    \\ \hline
2         &   60    &    90    \\ \hline
3         &   30    &    60    \\ \hline
4         &   15    &    30    \\ \hline
5         &   0    &     15   \\ \hline
\end{tabular}
\\
\caption{Definition of CKD stages. $\dagger$: eGFR values greater than 120 \unit are largely recognised as being inaccurate. For this reason 120 \unit is used as the cut-off between stages 0 and 1, despite the original KDIGO guideline~\cite{KDIGOGuidelines} not defining any upper value.}
\label{table:CKD_stage}
\end{table}

\subsection{The QICKD Dataset}
\label{data}
\begin{figure}
\setlength{\tabcolsep}{1pt}
\centering
\begin{tabular}{ccc}
\includegraphics[scale=0.22]{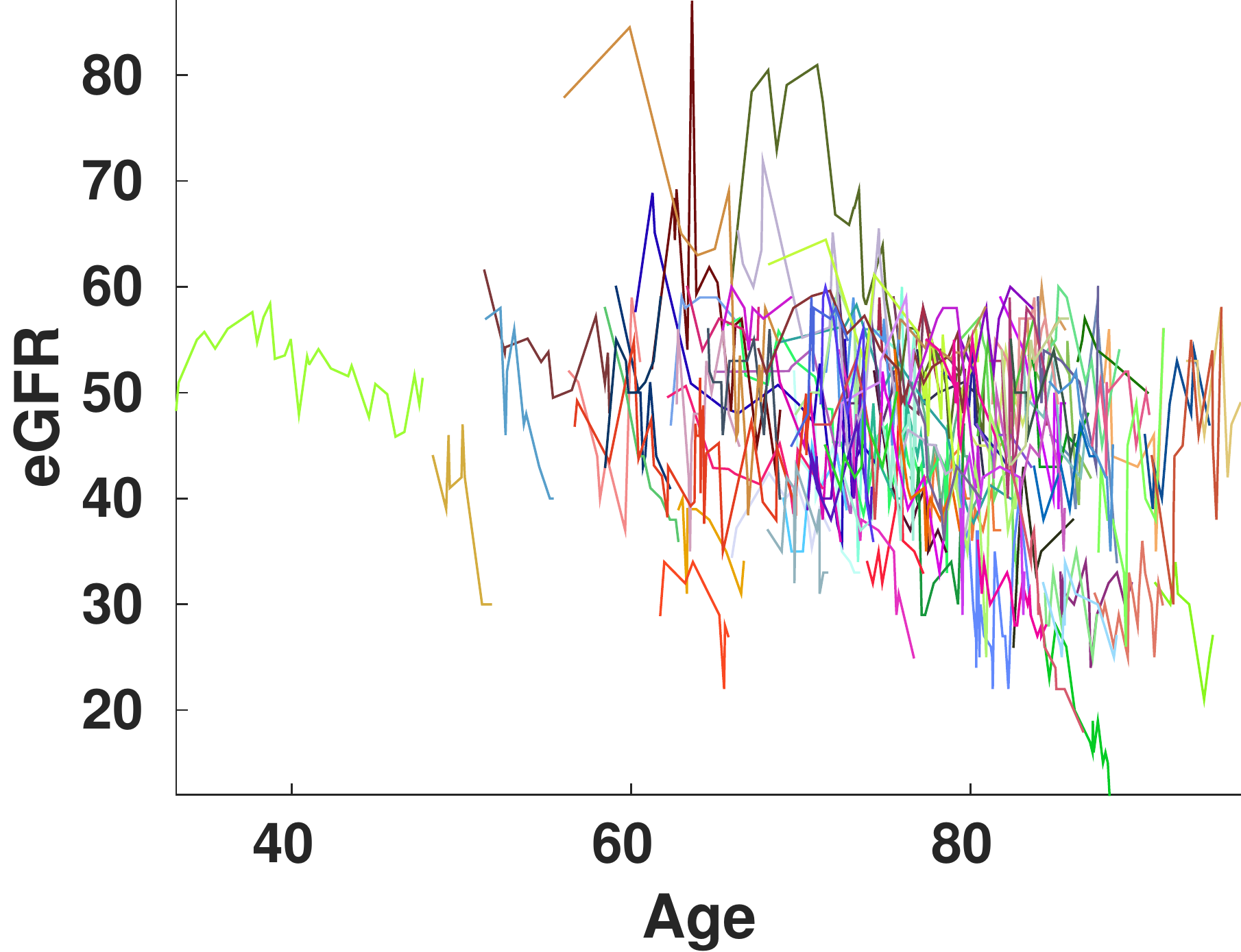} &
\includegraphics[scale=0.22]{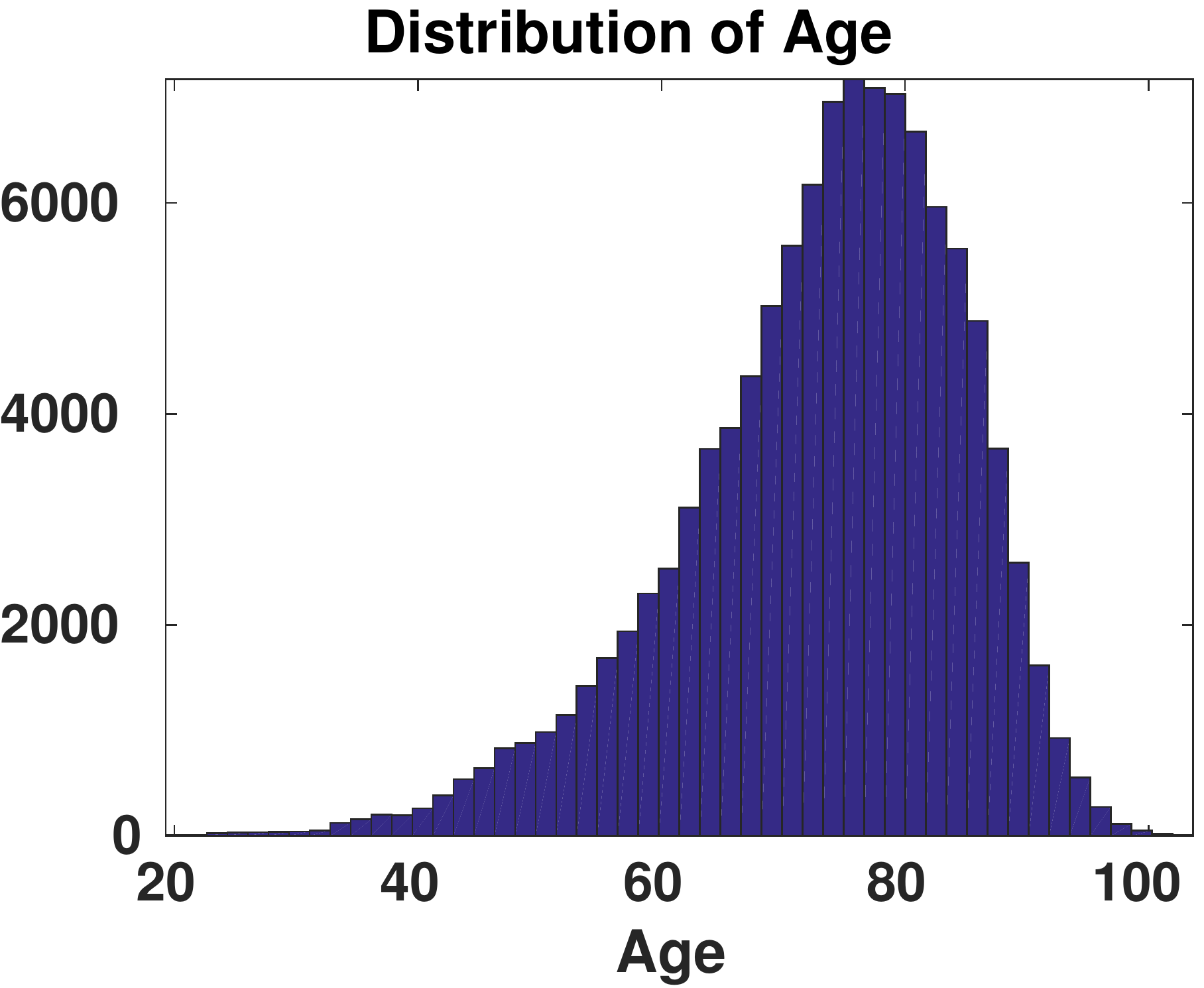} & \includegraphics[scale=0.22]{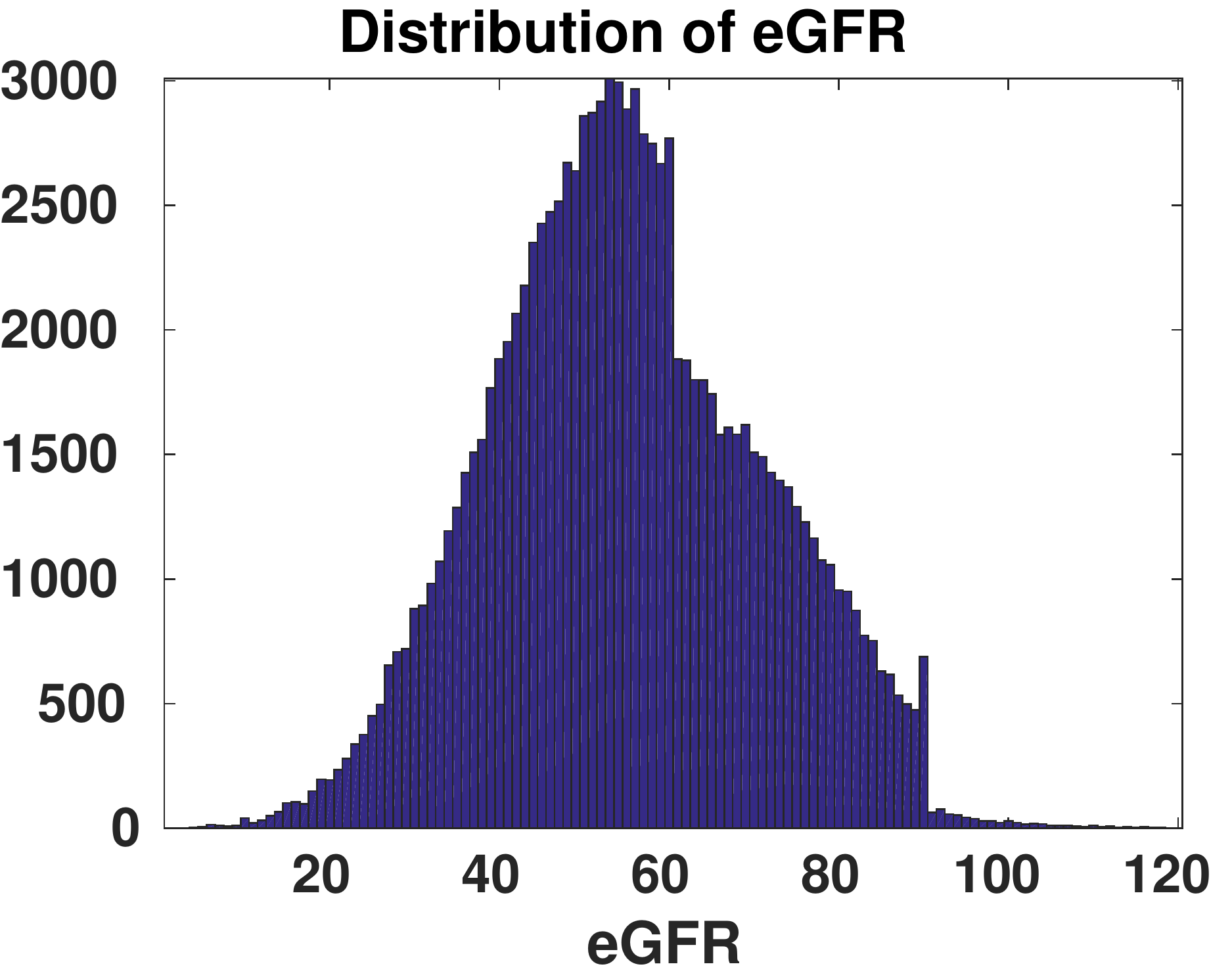} \\
(a) & (b) & (c) \\

 \includegraphics[scale=0.22]{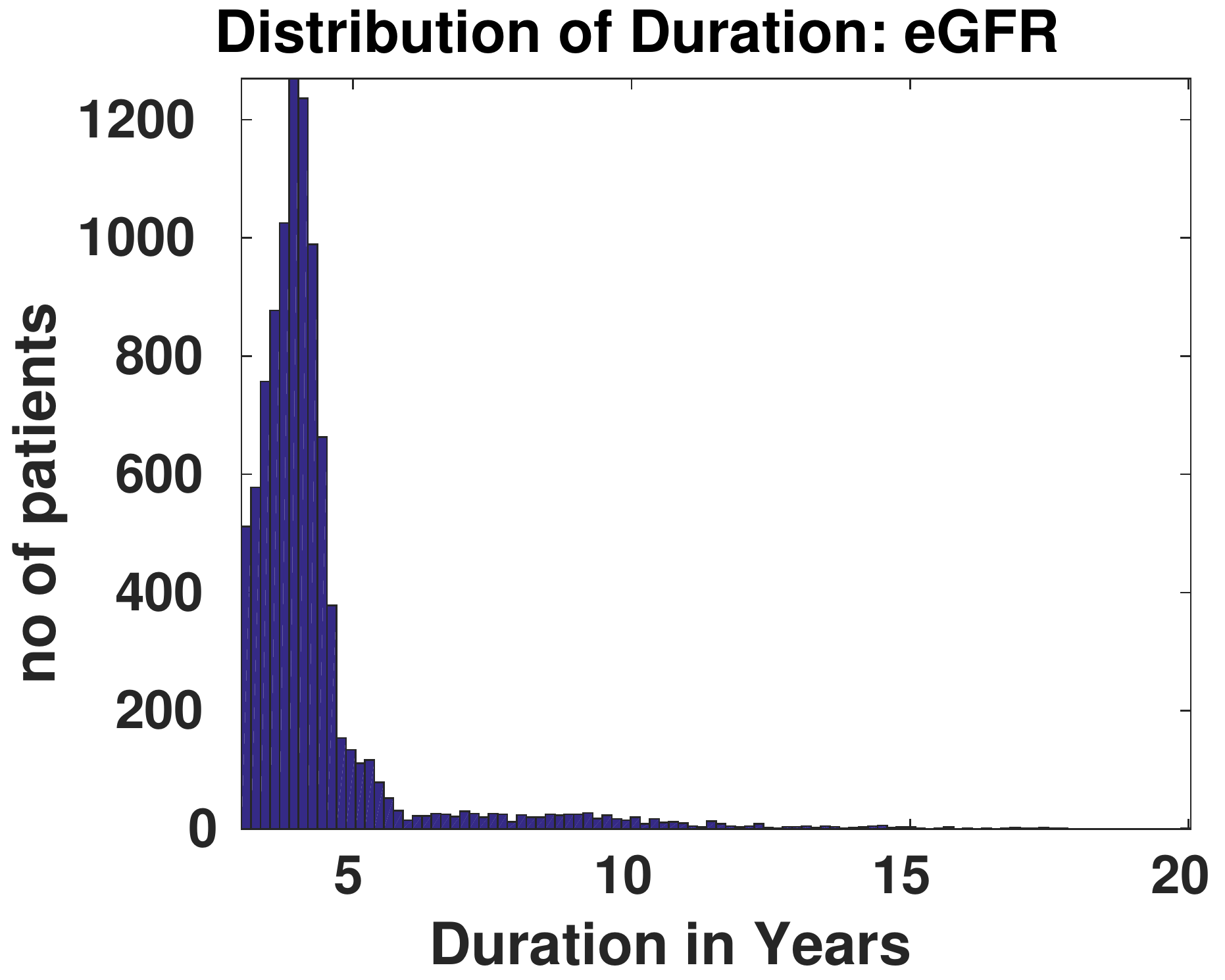} & \includegraphics[scale=0.22]{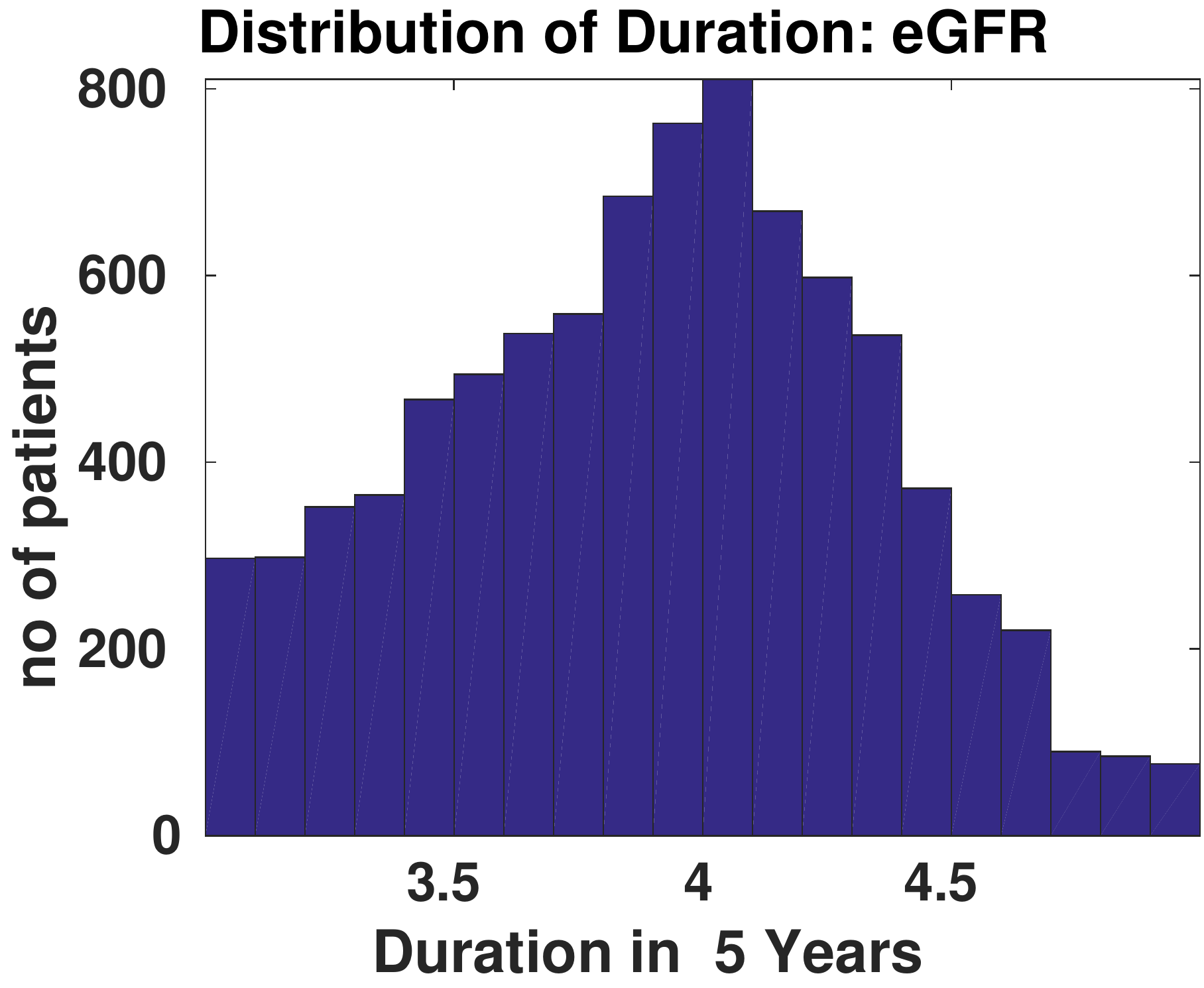} & \includegraphics[scale=0.22]{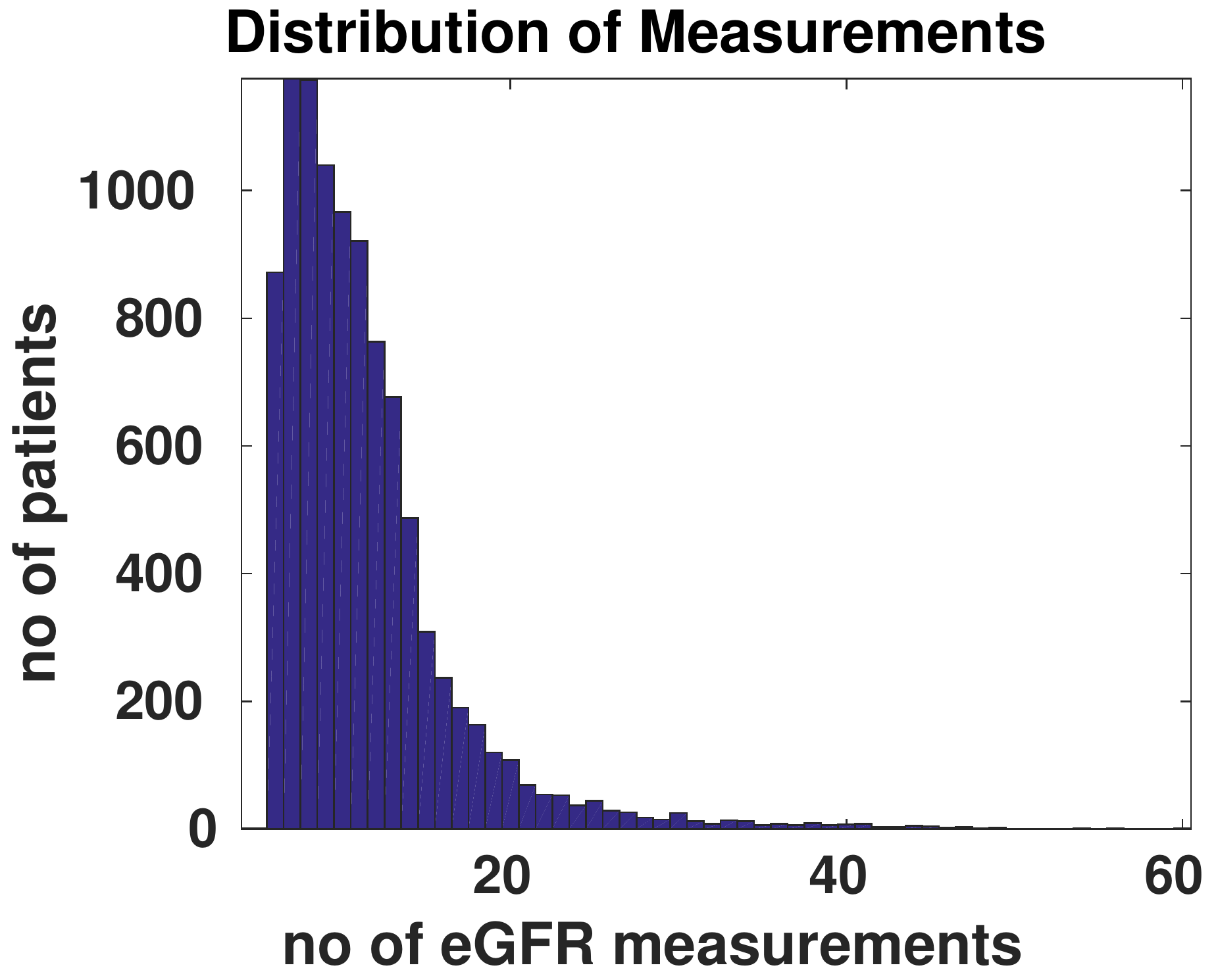} \\
(d) & (e) & (f) \\
\includegraphics[scale=0.22]{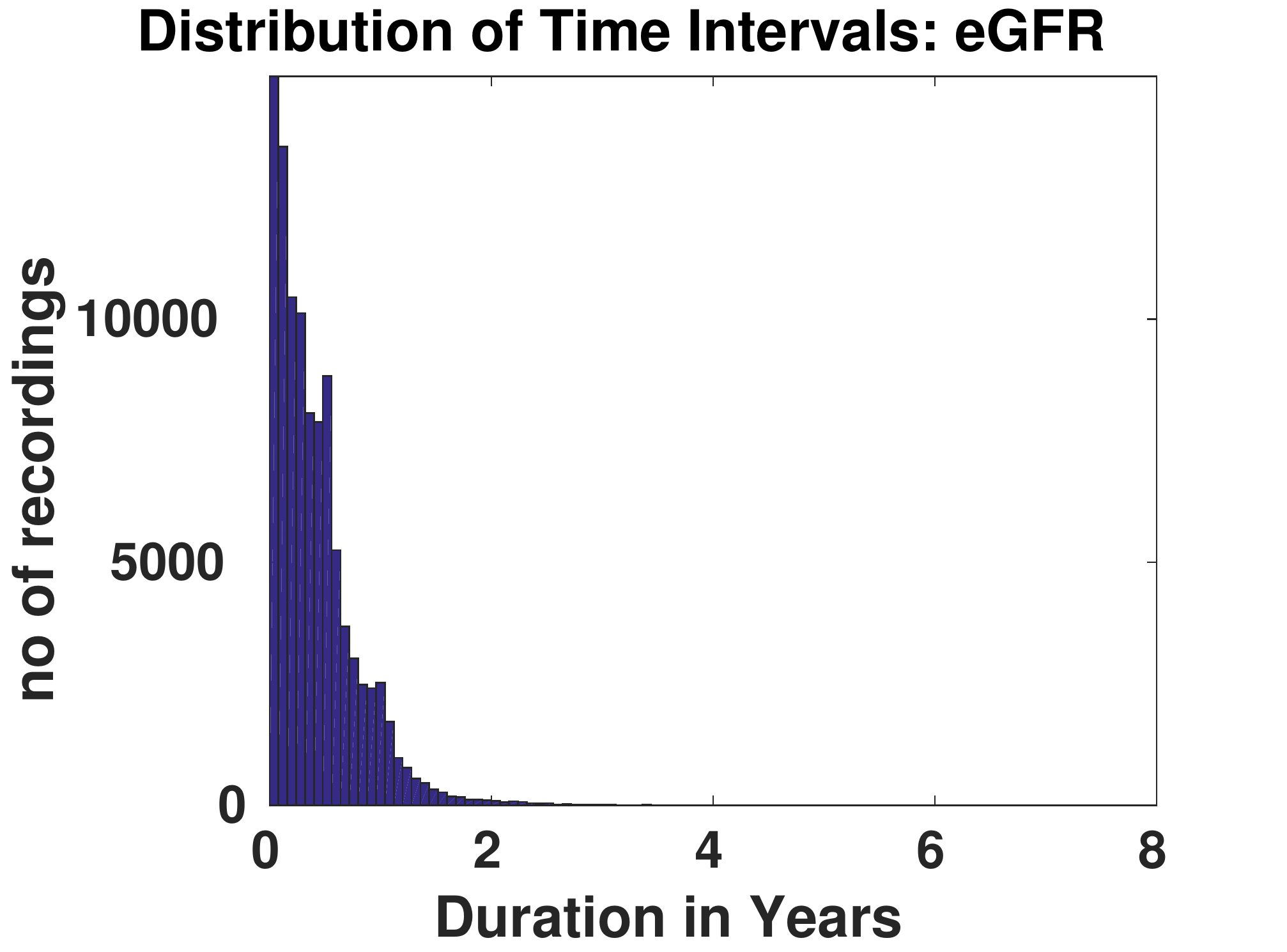} & \includegraphics[scale=0.22]{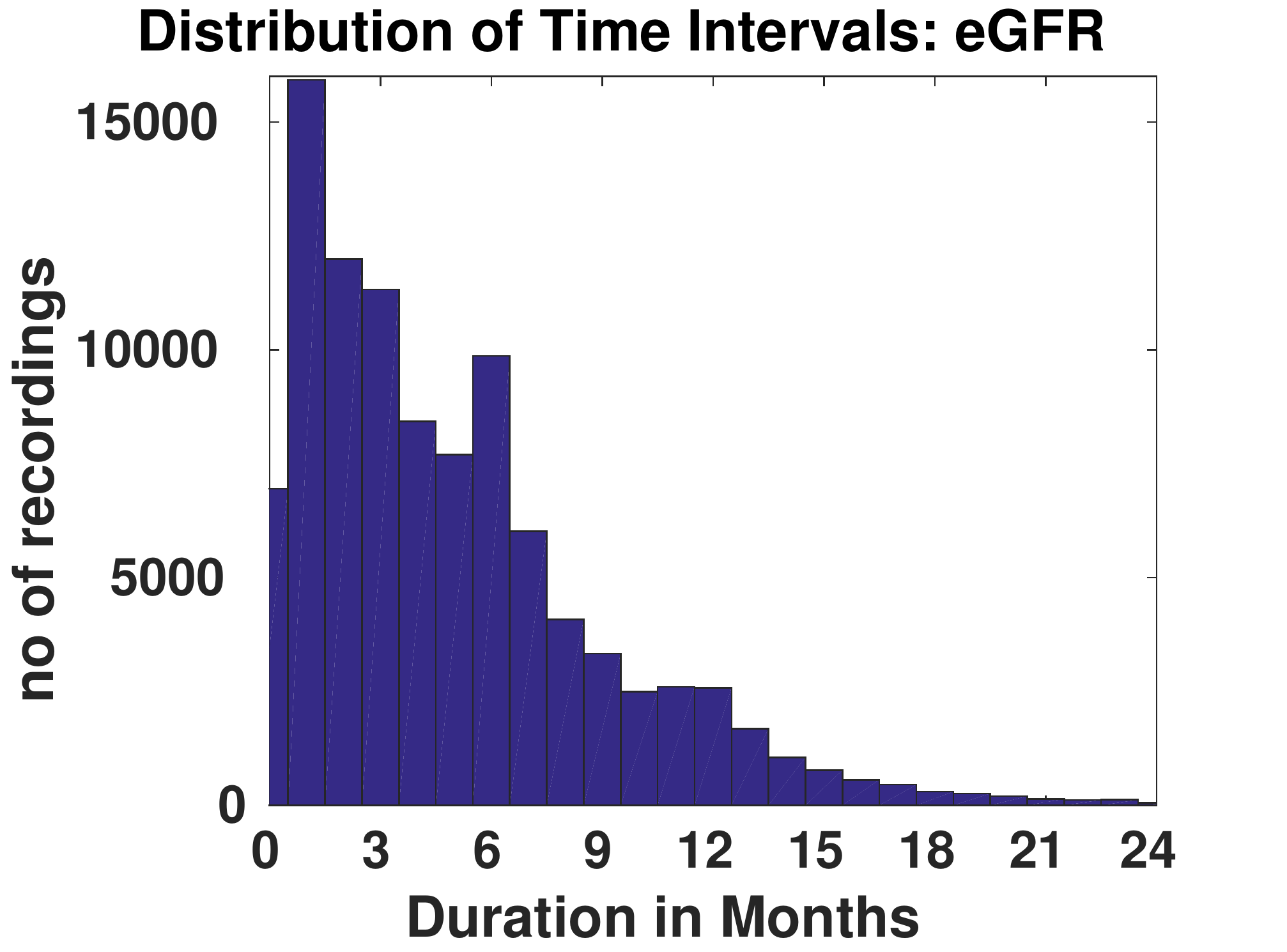} &  \\
(g) & (h) & \\
\end{tabular}
\caption{Patients' eGFR signals are observed at irregular time intervals and over different age ranges. (a) eGFR sequences for a subset of 100 patients (each colour represents a single patient). Dataset characteristics: distributions of the (b) ages over which patients had eGFR measurements recorded, (c) eGFR measurement values, (d) duration of time over which all of a patient's eGFR measurements were recorded, (e) the same as (d) but limited to those patients with all measurements occurring within five years, (f) number of eGFR measurements recorded per patient, (g) time intervals between consecutive eGFR measurements and (h) the same as (g) but limited to consecutive measurements occurring between one and twelve months apart.}
\label{fig:data}
\end{figure}

The QICKD dataset contains the primary care records of 951,764 patients. Of these records, 12,297 contain an eGFR measurement (45.4\% male and 56.6\% female). In total, there were 109,397 eGFR measurements across the patients, with approximately 95\% between the values of 25 and 120 \unit and occurring in patients between the ages of 60 and 103. Figure~\ref{fig:data} summarises the main characteristics of the dataset. Based on the patient statistics, a window length of three years and window interval of half a year were chosen as the most appropriate trade-off between smoothness and capturing local trends. Due to this, 1,546 of the 12,297 patients with an eGFR measurement were discarded for having less than three years worth of measurements and twenty-six for having gaps between measurements of larger than three years. Ten patients were also discarded for having an overly large gradient, likely as a result of large gaps between measurements isolating individual measurements. Each patient's eGFR sequence was also labelled, using the SAKIDA algorithm~\cite{tirunagari2016Detection}, with the number of acute kidney injury (AKI) episodes experienced by the patient. As an AKI represents a sudden and substantive change in the eGFR trend, and could therefore interfere with the trend modelling, the 1,103 patients identified as having experienced an AKI episode are excluded. In total 2,603 patients were excluded, leaving 9,694.

From Figure~\ref{fig:data}(c) it can be seen that there are abrupt changes in the distribution of the measurements at both 60 and 90 \unit . There are likely two reasons for this:
\begin{itemize}
\item Measurements above 60 or 90, depending on the testing laboratory, are truncated and reported as either 60 or 90 respectively. This practice results in a peak in the distribution at 60 and 90 \unit .
\item The reliability of eGFR measurements decreases as the value of the measurement increases due to the associated variance increasing with the value of the measurement. Therefore, eGFR measurements above certain values, 60 and 90, are often reported as greater than 60 or 90 but a numeric value is not recorded in the patient's record. This likely accounts for the observed drop in the distribution following 60 and 90 \unit .
\end{itemize}
Due to the increased variance of larger eGFR measurements, any values above 120 \unit are removed from the patient's time series and excluded from the study.

\subsection{Staging and Stratifying CKD}
Accurate staging of CKD is dependent on being able to take a stable measurement of a patient's eGFR. Therefore, one of the major drawbacks of current approaches to CKD staging is the variability of eGFR measurements due to their sensitivity to natural fluctuations in the breakdown of protein, e.g. from changing levels of protein in the diet, muscle breakdown and hydration, in addition to clinical factors. It is therefore possible for the trend in a patient's eGFR to be interrupted due to benign external factors, thereby masking a more serious decline in kidney function and frustrating clinicians' attempts to reliably determine a patient's CKD stage.

As an alternative to relying on the, potentially noisy, raw eGFR values when making staging decisions, here we use the estimated mean eGFR value obtained directly from the broken-stick model. To this end, we use a Bayesian framework to formulate the problem. Let $g_t$ be the eGFR measurement taken at time $t$ and $\{g_t|t\in\mathcal{T}\}$, where $\mathcal{T}\equiv \{1,\ldots, T \}$, be an eGFR time series. Then, after applying the broken-stick model, we have the corresponding regressed mean $\{\mu_t|t\in\mathcal{T}\}$. Using this notation, CKD stages can be determined using the following equation:
\begin{equation}
CKD(g) \equiv g_{\mathtt{L}}^{l} < g \le g_{\mathtt{U}}^{l} 
\end{equation}
where $g$ is a raw eGFR measurement value and $[g_{\mathtt{L}}^l, g_{\mathtt{U}}^l]$ is the eGFR range that defines a given CKD stage $l \in \{0,1,2,3,4,5\}$. The upper and lower bounds for each CKD stage are shown in Table~\ref{table:CKD_stage}. Given that there are no available ground-truth CKD stages, and as $\mu$ is an estimated eGFR value, we can perform the staging using it, rather than $g$, via the following equation :
\begin{equation}
CKD(\mu) \equiv g_{\mathtt{L}}^{l} < \mu \le g_{\mathtt{U}}^{l} 
\end{equation}

In order to ascertain whether the the CKD stages determined using $\mu$ and $g$ are consistent with one another, the distribution $p(g|CKD(\mu))$ was calculated for each CKD stage and the following equation used to determine the posterior probability of an expected given the raw eGFR value $g$:
\[
P(CKD(\mu)|g) = \frac {p(g|CKD(\mu)) P(CKD(\mu))} {\sum_{CKD(\mu^\dagger)} p(g|CKD(\mu^\dagger)) P(CKD(\mu^\dagger))}
\]
where $CKD(\mu)$ ranges from $0$ to $5$. In order to prevent the prior dominating the posterior, a uniform prior $P(CKD(\mu))$ was used, resulting in the following equation:
\begin{equation}
\label{eqn:posterior_CKD}
P(CKD(\mu)|g) = \frac {p(g|CKD(\mu))} {\sum_{CKD(\mu^\dagger)} p(g|CKD(\mu^\dagger))}
\end{equation}

\begin{figure}[tb]
  \centering
  \subfloat[Determining CKD stages]
  {\includegraphics[width=0.49\linewidth]{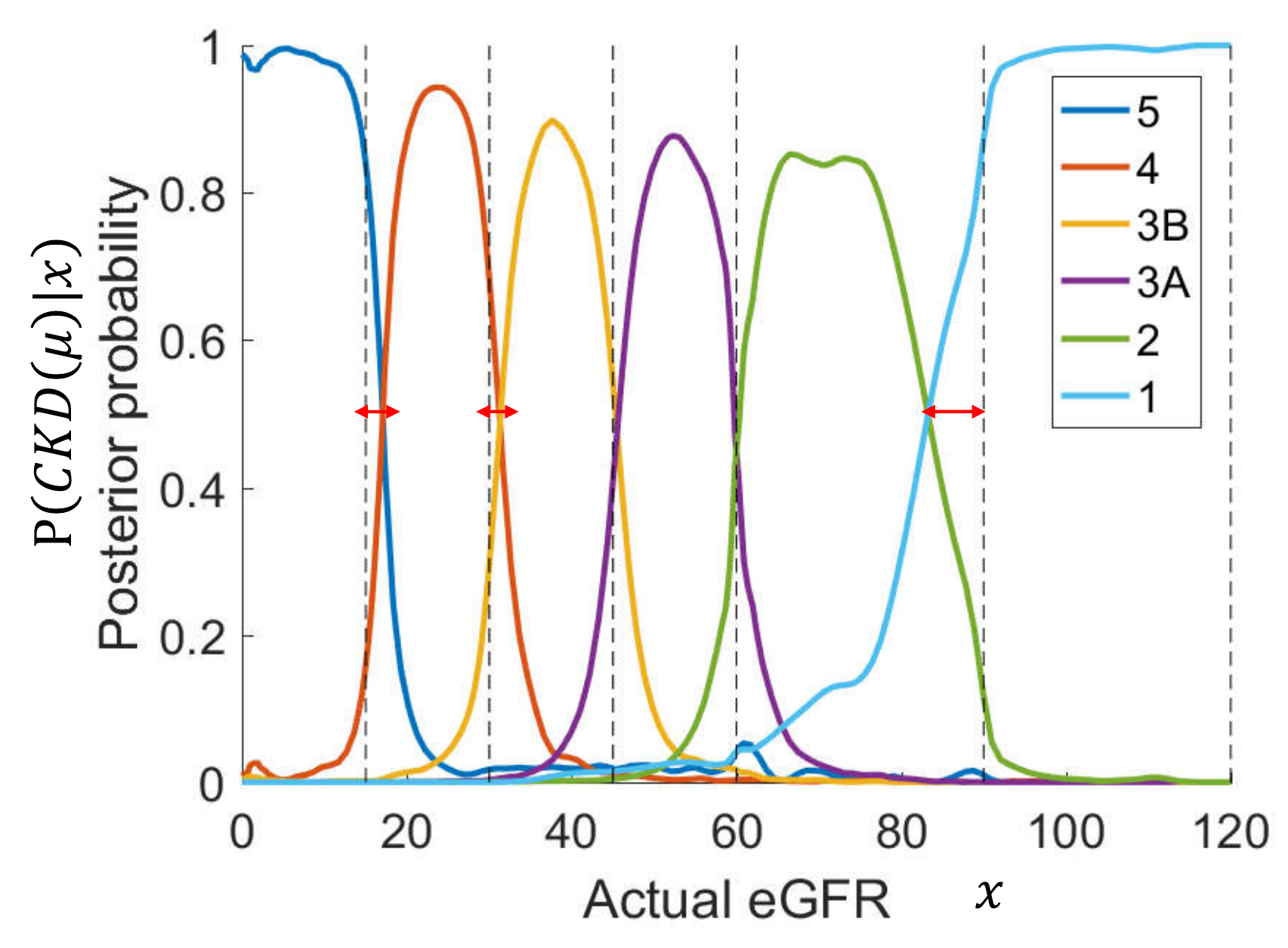}}
  \subfloat[Risk stratification using eGFR slope]
  {\includegraphics[width=0.49\linewidth]{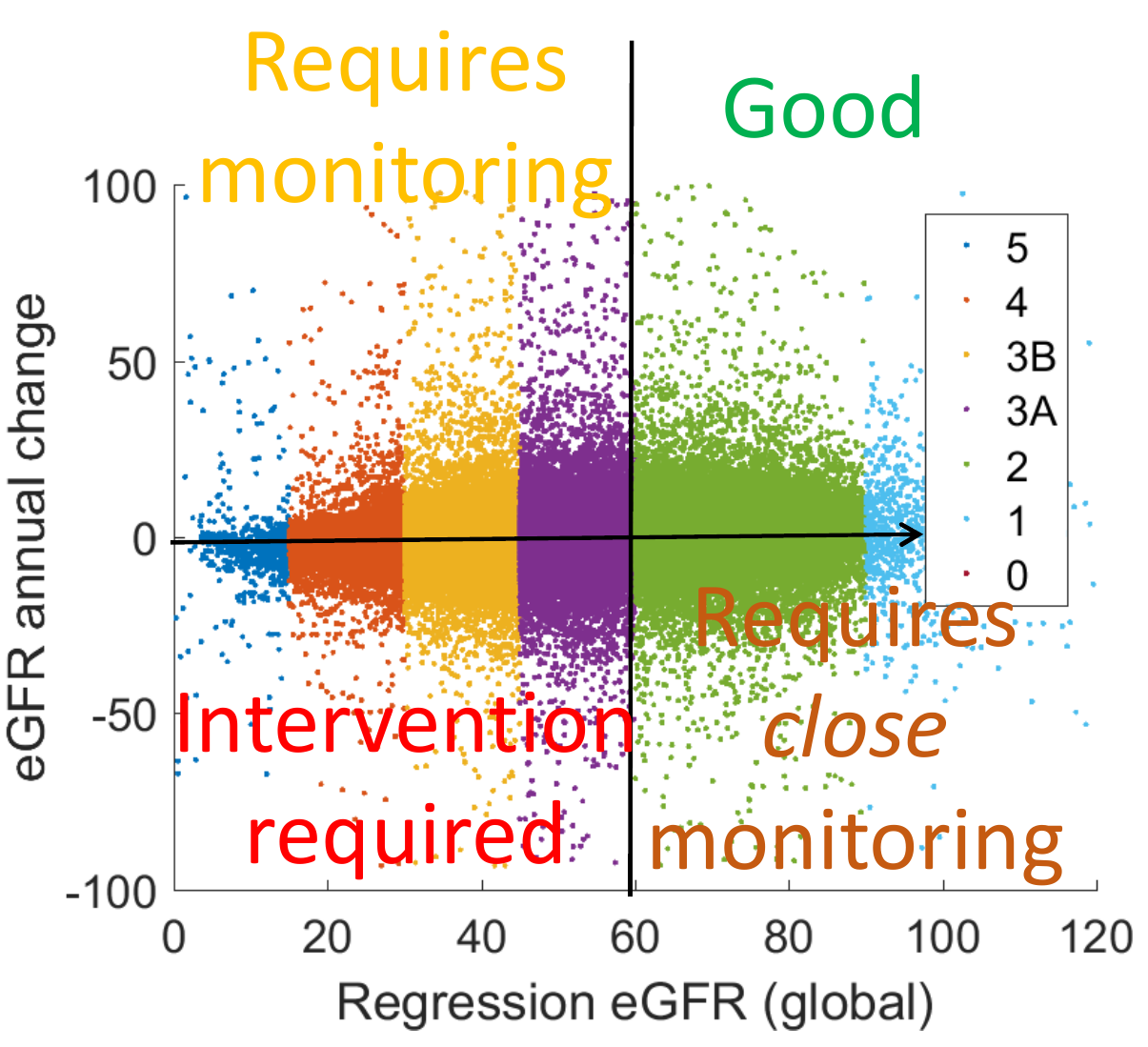}}
   \caption{\label{fig:app} Two applications of the broken-stick model. (a) Expected CKD stage posterior probabilities given raw eGFR values (the legend shows the expected CKD stage). Gaps between the KDIGO guideline stage (dashed vertical lines) and the boundary between expected CKD stages can be observed for stages 4/5, 3b/4 and 1/2. (b) Stratification of CKD patients using the expected eGFR slope (expected annual rate change). The use of the expected eGFR slope enables both the staging and trajectory of a patient's eGFR measurements to be taken into account.}
\end{figure}

The posterior probability distributions calculated using \Eqn{\ref{eqn:posterior_CKD}} can be seen in Figure~\ref{fig:app}(a). It is noticeable from this that the boundaries used to determine CKD stages from raw eGFR values, according to the KDIGO guidelines~\cite{KDIGOGuidelines} in Table~\ref{table:CKD_stage}, are not consistent with with the expected stage boundaries.

In addition to using the broken-stick model to determine CKD stages, by calculating the eGFR slope using \Eqn{\ref{eqn:rate_change}} it is possible to both stage and stratify patients according to the trajectory that their condition is taking. By calculating both the expected eGFR value ($\mu$) and slope ($\mu'$) at a given point, and recognising that stages 1 and 2 are often considered to be mild CKD, it is possible to stratify patients into four rough categories based on their outlook:
\begin{itemize}
\item Good: Patients in this category have mild, or no, CKD and a positive trajectory.
\item Requires monitoring: Patients in this category have more severe CKD but show a positive trajectory, and may therefore be less likely to have their CKD worsen.
\item Requires close monitoring: Patients in this category are characterised by having mild CKD but a worsening outlook due to the negative eGFR slope.
\item Intervention required: In this category patients have advanced and worsening CKD.
\end{itemize}

\section{Discussion and Conclusions}
The proposed broken-stick model can robustly estimate both short-term and long-term trends simultaneously, while also accommodating the unequal length and irregularly sampled nature of clinical time series. This can provide clinicians with a powerful tool for understanding the overall trajectory of a patient's disease progression by smoothing out local fluctuations in a parameterised manner. Within the management of CKD, the two primary uses of eGFR are determining the stage of a patient's CKD and determining the likely progression of the condition. While CKD staging is currently based on local trends, in this case the most recent eGFR measurements, by modelling a patient's eGFR time series using a broken-stick model it is possible to base a patient's stage on their entire time series. Conversely, evaluation of CKD progression can be based on both short- and long-term trends, and is difficult to evaluate from raw eGFR values alone. As a demonstration of the utility of the broken-stick model for assisting in the management of CKD, it was applied to the eGFR time series contained in the electronic medical records of approximately 10,000 patients. When compared to the CKD staging following the KDIGO guidelines, the stages determined using the broken-stick model are largely consistent, with the exception of between stages 1 and 2 where eGFR measurements are less reliable. Given this consistency, the gradient-based patient stratification is likely to prove reliable as it relies on the same model. Taken together, these results could provide useful information when determining the trajectory of a patient's condition and in the retrospective identification of patients for clinical research. Additionally, given its flexibility and wide applicability, the probabilistic broken-stick model could easily be applied to the modelling of additional biomedical measurements, such as plasma glucose concentration in diabetes.

\section*{Acknowledgements}
This work was supported by the Medical Research Council under grant number MR/M023281/1. We would like to thank our collaborators for providing us with the QICKD dataset. The project details can be found at \url{http://www.modellingckd.org/}.

\section*{References}
\bibliographystyle{model1-num-names}
\bibliography{santosh_references,AKI_detection,eGFRclass,BSM}






% \end{thebibliography}

\end{document}